\documentstyle[12pt]{article}

\newcommand{\bb}{\begin{equation}}
\newcommand{\ee}{\end{equation}}
\newcommand{\bqn}{\begin{eqnarray}}
\newcommand{\eqn}{\end{eqnarray}}
\newcommand{\pp}{\partial }

\newtheorem{theorem}{Theorem}

\begin{document}
\begin{titlepage}

\begin{flushright}

ULB--TH--96/11\\
hep-th/9606181\\
June 1996\\

\end{flushright}
\vfill

\begin{center}
{\Large{\bf  Characteristic cohomology of $p$-form gauge theories}}
\end{center}
\vfill

\begin{center}
{\large
Marc Henneaux$^{a,b}$, 
Bernard Knaepen$^{a,*}$ \\ and
Christiane Schomblond$^a$}
\end{center}
\vfill

\begin{center}{\sl
$^a$ Facult\'e des Sciences, Universit\'e Libre de Bruxelles,\\
Campus Plaine C.P. 231, B--1050 Bruxelles, Belgium\\[1.5ex]

$^b$ Centro de Estudios Cient\'\i ficos de Santiago,\\
Casilla 16443, Santiago 9, Chile

}\end{center}
\vfill

\begin{abstract}
The characteristic cohomology $H^k_{char}(d)$
for an arbitrary set of 
free $p$-form gauge fields 
is explicitly worked out in all form degrees $k<n-1$, where
$n$ is the spacetime dimension.  It is shown that
this cohomology is finite-dimensional
and completely generated by the forms dual to the field
strengths.  The gauge
invariant characteristic cohomology is also
computed.  The results are extended to interacting
$p$-form gauge theories with gauge invariant interactions.
Implications for the BRST cohomology
are mentioned.
\end{abstract}

\vspace{5em}

\hrule width 5.cm
\vspace*{.5em}

{\small \noindent (*)Aspirant du Fonds National de la
Recherche Scientifique (Belgium).}

\end{titlepage}

\section{Introduction}
\setcounter{equation}{0}
\setcounter{theorem}{0}

The characteristic cohomology \cite{Bryant}
plays a central role in the analysis of
any local field theory. The easiest way to define this
cohomology, which is contained in the so-called Vinogradov
$C$-spectral sequence \cite{Vinogr1,Vinogr2,Tsujishita}, 
is to start with the familiar notion of
conserved current.  Consider a dynamical theory with field
variables $\phi^i$ ($i=1,\dots ,M$) and Lagrangian ${\cal L}(\phi^i,
\pp_\mu \phi^i,\dots, \pp_{\mu_1 \dots \mu_k} \phi^i)$.  The field
equations read
\bb
{\cal L}_i = 0, \label{FE0}
\ee
with
\bb
{\cal L}_i = \frac{\delta {\cal L}}{\delta \phi^i} =
\frac{\pp {\cal L}}{\pp \phi^i} - \pp_\mu
\big( \frac{\pp {\cal L}}{\pp (\pp_\mu \phi^i)} \big) +
\dots + (-1)^k \pp_{\mu_1 \dots \mu_k} \big( \frac{\pp {\cal L}}{\pp 
(\pp_{\mu_1 \dots \mu_k} \phi^i)} \big).
\ee
A (local) conserved current $j^\mu$ is a vector-density which
involves the fields and their derivatives up to some
finite order and which is conserved modulo the field 
equations, i.e., which fulfills
\bb
\pp_\mu j^\mu \approx 0.\label{ConsCurr}
\ee 
Here and in the sequel, $\approx $ means ``equal 
when the equations of motion hold" or, as one also says
equal ``on-shell".  Thus, (\ref{ConsCurr}) is equivalent to
\bb
\pp_\mu j^\mu = \lambda^i {\cal L}_i + \lambda^{i \mu}
\pp_\mu {\cal L}_i + \dots + \lambda^{i \mu_1 \dots \mu_s}
\pp_{\mu_1 \dots \mu_s} {\cal L}_i
\label{ConsCurr2}
\ee
for some $\lambda^{i \mu_1 \dots \mu_j}$, $j=0,\dots,s$.
A conserved current is said to be trivial if it can be written as
\bb
j^\mu \approx \pp _\nu S^{\mu \nu} \label{TrivCurr} 
\ee
for some local antisymmetric tensor density $S^{\mu \nu} = -
S^{\nu \mu}$.  The terminology does not mean that trivial
currents are devoid of physical interest, but rather, that they
are easy to construct and that they are trivially conserved. 
Two conserved currents are said to be equivalent if they differ
by a trivial one.  The characteristic cohomology in degree $n-1$ is
defined to be the quotient space of equivalence classes of
conserved currents.  One assigns the degree $n-1$ because
the equations (\ref{ConsCurr}) and (\ref{TrivCurr}) can be
rewritten as
$d \omega \approx 0$
and
$\omega \approx d \psi$
in terms of the ($n-1$)-form $\omega$ and ($n-2$)-form
$\psi$ respectively dual to $j^\mu$ and $S^{\mu \nu}$.

One defines the characteristic cohomology in degree $k$
($k<n$) along exactly the same lines, by simply considering other
values of the form degree.  So, one says that a local
$k$-form $\omega$ is a cocycle of the characteristic 
cohomology in degree $k$ if it is weakly closed,
\bb
d \omega \approx 0; \;\; \hbox{``cocycle condition"}
\label{cocycleCC}
\ee
and that it is a coboundary if it is weakly exact,
\bb
\omega\approx d \psi, \;\; \hbox{``coboundary condition"}
\label{coboundaryCC}
\ee
just as it is done for $k=n-1$.
For instance, the characteristic cohomology in form degree
$n-2$ is defined, in dual notations, as the quotient space of
equivalence classes of weakly conserved antisymmetric tensors,
\bb
\pp _\nu S^{\mu \nu} \approx 0, \; S^{\mu \nu} = S^{[\mu \nu]},
\label{ConsLaw}
\ee
where two such tensors are regarded as equivalent iff
\bb
S^{\mu \nu} - S^{'\mu \nu} \approx  \pp_\rho R^{\rho\mu \nu}, \; 
R^{\rho\mu \nu} = R^{[\rho\mu \nu]}.  
\ee
We shall denote the characteristic cohomological 
groups by $H^k_{char}(d)$.

Higher order conservation laws 
involving antisymmetric tensors
of degree $2$ or higher 
are quite interesting in their own right. 
In particular, conservation laws of the form (\ref{ConsLaw}), 
involving an antisymmetric tensor $S^{\mu \nu}$ 
have attracted a great deal of interest in the past
\cite{Unruh} as well as recently \cite{BBH3,Torre} in the context of
the mechanism of ``charge without charge" of Wheeler 
\cite{MisnerWheeler}.

But the characteristic cohomology is also important for
another reason: it appears as an  
auxiliary cohomology in the calculation of the local BRST
cohomology \cite{BBH1}.  This local BRST cohomology, in turn,
is quite useful in the determination of the structure of the
counterterms \cite{Tyutin,GomisWeinberg} and the anomalies
\cite{Stora} in the quantum theory. It plays also a central role
classically, in constraining the form of the consistent
deformations of the action
\cite{BarnichHenn}.  It is by establishing
vanishing theorems for the characteristic cohomology 
that the problem of consistent deformations and of candidate
anomalies has been completely solved in the 
cases of Yang-Mills gauge theories
and of gravity
\cite{BBH2,BBH3}. 
For this reason, it is an important question to determine the
characteristic cohomological groups 
for any given theory.

The purpose of this paper is to carry out this task for a system
of free antisymmetric tensor fields $B^a_{\mu_1 \dots \mu_{p_a}}$, 
$a= 1, \dots, N$, with Lagrangian
\bb
{\cal L} = \sum_a \big({-1 \over 2(p_a +1)!}H^a_{\mu_1 \ldots 
\mu_{p_a+1}} H^{a \mu_1
\ldots \mu_{p_a+1}}\big) \label{Lagrangian} 
\ee 
where the $H^a$'s are the ``field strengths" or ``curvatures",
\bqn
H^a&=&{1\over (p_a+1)!} H^a_{\mu_1 \ldots \mu_{p_a+1}}dx^{\mu_1} 
\ldots dx^{\mu_{p_a+1}} =dB^a \label{FieldStrength}\\
B^a&=&{1\over p_a!} B^a_{\mu_1 \ldots \mu_{p_a}} dx^{\mu_1} \ldots
dx^{\mu_{p_a}}
\eqn
The equations of motion, obtained by varying the fields
$B^a_{\mu_1 \ldots \mu_{p_a}}$, are given by
\begin{equation}
\partial_{\rho} H^{a\rho \mu_1 \ldots \mu_{p_a}} =0. \label{FE1}
\end{equation}
We consider simultaneously antisymmetric tensors  of different
degrees, but we assume $1 \leq p_a$.
We also assume $n>p_a + 1$ for each $a$ so that the
fields $B^a_{\mu_1 \dots \mu_{p_a}}$ all carry local degrees
of freedom.  Modifications of the Lagrangian by gauge invariant
interactions are treated at the end of the paper.

We give complete results for the characteristic cohomology in degree
$<n-1$, that is, we determine all the
solutions to the equation $\pp_{\mu} S^{\mu \nu_1 \dots \nu_s} 
\approx 0$ with $s>0$. 
Although we do not solve the characteristic cohomology
in degree $n-1$, we comment on the gauge invariance properties
of the conserved currents
and provide an infinite number of them,
generalizing earlier results of the Maxwell case
\cite{Lipkin,Morgan,ZZZ}\footnote{The 
determination
of all the conserved currents is of course also an interesting
question, but it is not systematically 
pursued here for two reasons.  First
the characteristic cohomology $H^{n-1}_{char}(d)$ is 
infinite-dimensional for the free theories considered
here and does not appear to be completely known even in the Maxwell
case in an arbitrary number of dimensions. 
By contrast, the cohomological groups
$H^{k}_{char}(d)$, $k<n-1$, are all finite-dimensional
and can be explicitly computed. 
Second, 
the group
$H^{n-1}_{char}(d)$ plays no role in the analysis of the consistent
interactions of antisymmetric tensor fields of degree $>1$,
as well as in the analysis of candidate anomalies if the
antisymmetric tensor fields all have degree $>2$ \cite{HKS}.}.
The
results of this paper will be used in \cite{HKS}
to compute the BRST cohomology of free, antisymmetric tensor fields.
This is a necessary step not only for determining the
possible consistent  interactions that can be added to the free
Lagrangian, but also for analyzing completely the BRST
cohomology in the interacting case.  Our results
have already been used and partly announced in \cite{MH} to show
the uniqueness of the Freedman-Townsend deformation of
the gauge symmetries of a system 
of antisymmetric tensors of degree $2$ in four dimensions.

Antisymmetric tensor fields - or, as one also says, $p$-form gauge 
fields - have been much studied in the past
\cite{Ogiev,Kalb,Freedman,Nepom,Teitel} and are crucial ingredients of
string theory and of various supergravity models \cite{GSW}.
The main feature of theories involving $p$-form gauge fields
is that their gauge symmetries
are {\em reducible}.  More precisely, in the present case, the 
Lagrangian (\ref{Lagrangian}) is invariant under the gauge
transformations
\bb
B^a \rightarrow B^{'a} = B^a + d\Lambda^a  \label{GaugeSym}
\label{bkgaugeinv}
\ee
where $\Lambda^a$ are arbitrary ($p_a-1$)-forms.  
Now, if $\Lambda^a = d\epsilon^a$,
then, the variation of $B^a$ vanishes identically.  Thus, the gauge 
parameters
$\Lambda^a$ do not all provide independent gauge symmetries:  
the gauge transformations (\ref{GaugeSym}) are reducible.
In the same way, if $\epsilon^a$ is equal to $d\mu^a$,
then, it yields a vanishing $\Lambda^a$.
There is ``reducibility of reducibility"
unless $\epsilon^a$ is a zero form.  If $\epsilon^a$ is not
a zero form, 
the process keeps going until one reaches $0$-forms. For 
the theory with Lagrangian (\ref{Lagrangian}), there
are thus 
$p_M-1$ stages of reducibility of the
gauge transformations ($\Lambda^a$ is a ($p_a-1$)-form),
where $p_M$ is the
degree of the form of higher degree occurring in (\ref{Lagrangian})
\cite{Siegel,Thierry1,Thierry2,Baulieu1}. 
One says 
that the  theory
is a  reducible gauge theory of reducibility order
$p_M-1$.
 
General vanishing theorems have been established in
\cite{Bryant,Vinogr1,Vinogr2,BBH1} showing that the characteristic 
cohomology of reducible theories of reducibility order
$p-1$  vanishes in form degree
strictly smaller than $n-p-1$.  Accordingly,  in the case of
$p$-form gauge theories, 
there can be a priori non-vanishing
characteristic cohomology only in form degree $n-p_M-1$, $n-p_M$, etc,
up to form degree $n-1$ (conserved currents). 
In the $1$-form case,
these are the best vanishing theorems one can prove, since a set of
free gauge fields $A^a_\mu$ has characteristic cohomology
both in form degree $n-1$ and $n-2$ \cite{BBH1}. Representatives
of the cohomology classes in form degree $n-2$ are given
by the duals to the field strengths, which are indeed closed
on-shell due to Maxwell equations.

Our main result is that the general vanishing theorems of
\cite{Bryant,Vinogr1,Vinogr2,BBH1} can be considerably
strengthened when $p >1$. 
For instance, if there
is a single $p$-form gauge field and if $n-p-1$ is odd,
there is only one non-vanishing group of
the characteristic cohomology in degree $<n-1$.  This is 
$H^{n-p-1}_{char} (d)$, which is one-dimensional. 
All the other groups 
$H^{k}_{char}(d)$ of the
characteristic cohomology with  $n-p-1<k<n-1$ are
zero, even though the general theorems of 
\cite{Bryant,Vinogr1,Vinogr2,BBH1}
leave open the possibility that they do not vanish. As we shall show
in \cite{HKS}, it is the presence of these additional
zeros that give $p$-form gauge fields and gauge transformations
their strong rigidity. 
 
Besides the standard characteristic cohomology, one may consider
the invariant characteristic cohomology, in which the local forms
$\omega$ and $\psi$
occurring in (\ref{cocycleCC}) and (\ref{coboundaryCC}) are
required to be invariant under the
gauge transformations (\ref{GaugeSym}).  We also
completely determine in this paper the invariant characteristic
cohomology in form degree $<n-1$.

Our method for
computing the characteristic cohomology is based
on the reformulation performed in \cite{BBH1}
of the characteristic cohomology
in form degree $k$ in terms of the cohomology $H^n_{n-k}(\delta|d)$
of the Koszul-Tate differential
$\delta$ modulo the spacetime exterior derivative $d$.
Here, $n$ is the form degree and
$n-k$ is the antighost number.  This approach is strongly
motivated by the BRST construction and appears to be 
particularly attractive and powerful.

Our paper is organized as follows.
In the next section, we formulate precisely our main results,
which are (i) that the characteristic cohomology $H^{k}_{char}(d)$
with  $k<n-1$ 
is generated (in the exterior product) by the exterior forms
${\overline H}^{a}$ dual to the field strengths $H^a$; these are 
forms of
degree $n-p_a-1$; and (ii) that the invariant characteristic
cohomology $H^{k, inv}_{char}(d)$ with  $k<n-1$
is generated (again in the exterior product) by the exterior forms
$H^a$  and ${\overline H}^{a}$.
We then review, in sections 3 and 4,
the definition and properties of the Koszul-Tate complex. 
Section 5 is of a more technical nature and relates
the characteristic cohomology to the cohomology
of the differential $\delta +d$, where $\delta$ is the
Koszul-Tate differential.  Section 6 analyses the
gauge invariance properties of $\delta$-boundaries modulo $d$. 
In section 7, we determine the characteristic 
cohomology for a single
$p$-form gauge field.
The results are then extended to an arbitrary system of $p$-form
gauge fields in section 8.  The invariant cohomology is analyzed 
in section 9.  Section 10 discusses in detail the 
cohomological groups $H^*(\delta \vert d)$, which play a key role in
the calculation of the local BRST cohomological groups 
$H^*(s \vert d)$.In section 11, we show that the existence of
representatives  expressible in terms of the ${\overline H}^{a}$'s
does not extend  
to the characteristic cohomology in form degree $n-1$,
by exhibiting an infinite number of (inequivalent)
conserved currents which are not of that form.  We
show next in section 12 that the results on the free
characteristic cohomology in degree $<n-1$ 
can be generalized straightforwardly
if one adds to the free Lagrangian (\ref{Lagrangian})
gauge invariant interaction terms that involve the fields 
$B^a_{\mu_1 \dots \mu_{p_a}}$ and their derivatives only
through the gauge invariant field strength components
and their derivatives (which are in general the only
consistent interactions that one can add).
We conclude in section 13 by summarizing our results and 
indicating future lines of research.

We assume throughout this paper that spacetime is the
n-dimensional Minkowski space, so that the indices in
(\ref{Lagrangian}) are raised with the inverse $\eta^{\mu\nu}$ of
the flat Minkowski metric $\eta_{\mu\nu}$. However, because of their
geometrical character, our results generalize
straightforwardly to curved backgrounds.

\section{Results}
\setcounter{equation}{0}
\setcounter{theorem}{0}

\subsection{Characteristic cohomology}

The equations of motion  (\ref{FE1}) can be rewritten as
\bb
d {\overline H}^a \approx 0  \label{FE2}
\ee
in terms of the ($n-p_a-1)$-forms ${\overline H}^a$
dual to the field strengths.  It then follows that any polynomial
in the ${\overline H}^a$'s is closed on-shell and thus defines
a cocycle of the characteristic cohomology.

The remarkable feature is that these polynomials are
not only inequivalent in cohomology, but
also {\em completely
exhaust the characteristic cohomology in form degree
strictly smaller than $n-1$}.  Indeed, one has:
\begin{theorem}
Let ${\cal {\overline H}}$ be the algebra generated by
the ${\overline H}^a$'s and let ${\cal V}$ be the
subspace containing the polynomials in the ${\overline H}^a$'s
with no term of form degree exceeding $n-2$.
The subspace ${\cal V}$ is isomorphic to
the characteristic cohomology in form degree
$<n-1$.
\label{MainResult}
\end{theorem}
We stress
again that the theorem does not hold in degree $n-1$ 
because there exist conserved
currents not expressible in terms of the ${\overline H}^a$'s.

Since the form degree is limited by the spacetime dimension
$n$, and since ${\overline H}^a$ 
has form degree
$n-p_a-1$, which is strictly positive (as explained in
the introduction, we assume $n-p_a-1>0$ for each $a$),
the algebra ${\cal {\overline H}}$ 
is finite-dimensional.  In that algebra, the
${\overline H}^a$ with even $n-p_a-1$ commute with all the
other generators, while the ${\overline H}^a$ with odd $n-p_a-1$
are anticommuting objects.

\subsection{Invariant characteristic cohomology}
While the cocycles of Theorem {\bf 2.1} are all gauge invariant,
there exists co\-boundaries of the characteristic cohomology
that are gauge invariant, i.e., that involve only the field
strength components and their derivatives, but which cannot,
nevertheless, be written as coboundaries of  gauge invariant
local forms, even weakly. Examples are given by the field
strengths $H^a = dB^a$ themselves. For this reason, the
invariant characteristic cohomology and the characteristic
cohomology do not coincide. We shall denote by
${\cal H}$ the finite-dimensional algebra generated by the
$(p_a+1)$-forms $H^a$, and by ${\cal J}$ the finite-dimensional
algebra generated by the field strengths $H^a$ and their
duals ${\overline H}^a$.  One has
 
\begin{theorem}
Let ${\cal W}$ be the
subspace of  ${\cal J}$  containing the polynomials
in the $H^a$'s and the ${\overline H}^a$'s
with no term of form degree exceeding $n-2$.
The subspace ${\cal W}$ is isomorphic to
the invariant  characteristic cohomology in form degree
$<n-1$.
\label{MainResult2}
\end{theorem}
Our paper is devoted to proving these theorems. 

\subsection{Cohomologies in algebra of $x$-independent forms}

The previous theorems hold as they are formulated
in the algebra of local forms
that are allowed to have an explicit $x$-dependence.  The explicit
$x$-dependence enables one to remove the constant $k$-forms ($k>0$)
from the cohomology, since these are exact, $c_{i_1 i_2 \dots
i_k} dx^{i_1} dx^{i_2} \dots dx^{i_k} = d(c_{i_1 i_2 \dots
i_k} x^{i_1}$ $dx^{i_2} \dots dx^{i_k})$.  If one restricts one's
attention to the algebra of local forms with no explicit dependence
on the spacetime coordinates, then, one must replace in the
above theorems the polynomials in the curvatures and their duals
with coefficients that are {\em numbers} by 
the polynomials in the curvatures
and their duals with coefficients that are {\em constant exterior
forms}.

Note that the constant exterior forms 
can be alternatively gotten rid of without introducing
an explicit $x$-dependence, by imposing
Lorentz invariance (there is no Lorentz-invariant
constant $k$-form for $0<k<n$). 

\section{Koszul-Tate Complex}
\setcounter{equation}{0}
\setcounter{theorem}{0}

The definition of the cocycles of the characteristic co\-homology
$H^k_{char}(d)$ involves ``weak" equa\-tions holding only
on-shell.  It is convenient to replace them by ``strong"
equations holding everywhere in field space, and not just when
the equations of motion are satisfied.
The reason is that the coefficients
of the equations of motion in the conservation laws are not
arbitrary, but are subject to restrictions whose analysis
yields useful insight on the conservation laws themselves. From
this point of view, the
equation (\ref{ConsCurr2}) involving the
coefficients $\lambda^{i \mu_1 \dots \mu_j}$ is a more
interesting starting point than Eq. (\ref{ConsCurr}).
One useful  way to replace weak equations
by strong equations is to introduce the Koszul-Tate resolution
associated with the equations of motion (\ref{FE1}).

The details of the construction of the Koszul-Tate differential
$\delta$ can be found in \cite{HenneauxTeitelboim}. Because the
present theory is reducible, we must introduce the 
following set of  BV-antifields \cite{BV}:
\begin{equation}
B^{*a \mu_1 \ldots \mu_{p_a}},
B^{*a\mu_1 \ldots \mu_{p_a-1}},\ldots,
B^{*a\mu_1},B^{*a}.
\label{antifieldlist}
\end{equation}
The Grassmann parity and the {\it antighost} number of 
the antifields $B^{*a \mu_1
\ldots \mu_{p_a}}$ associated with
the fields $B^a_{\mu_1 \ldots \mu_{p_a}}$ are equal to $1$. 
The Grassmann parity and the {\it antighost} number of the
other antifields is determined according to the
following rule.  As one moves from one
term to the next one to its right in (\ref{antifieldlist}), 
the Grassmann parity
changes and the antighost number increases by one unit.
Therefore the parity and the
antighost number of a given antifield
$B^{*a \mu_1 \ldots \mu_{p-j}}$ are respectively $j+1$ modulo $2$
and $j+1$.
 
The Koszul-Tate differential acts in the 
algebra ${\cal P}$ of local exterior
forms.  By definition, a local exterior form $\omega$
reads
\bb
\omega = \sum \omega_{\mu_1 \dots \mu_J} dx^{\mu_1} \dots 
dx^{\mu_J}
\label{LocalForm}
\ee
where the coefficients $\omega_{\mu_1 \dots \mu_J}$
are smooth functions of the coordinates $x^\mu$, the fields 
$B^a_{\mu_1 \ldots \mu_{p_a}}$, the antifields 
(\ref{antifieldlist}), and their
derivatives up to a finite order.  Although this is not
strictly necessary, we shall actually
assume polynomiality in the fields, the antifields and their
derivatives, as this is the situation encountered in field
theory. 

The Koszul-Tate differential is defined by its action on the
fields and the antifields as follows:

\begin{eqnarray}
\delta B^a_{\mu_1 \ldots \mu_{p_a}}&=&0, 
\label{DefKT0}\\
\delta B^{*a \mu_1 \ldots \mu_{p_a}}&=&\partial_{\rho} H^{a \rho \mu_1
\ldots\mu_{p_a}}, 
\label{DefKT1}\\
\delta B^{*a\mu_1 \ldots \mu_{p_a-1}}&=&\partial_\rho B^{*a \rho \mu_1
\ldots \mu_{p_a-1}}, \label{DefKT2}\\ \nonumber &\vdots& \\
\delta B^{*a\mu_1}&=&\partial_\rho B^{*a\rho \mu_1}, 
\label{DefKT3}\\
\delta B^{a*} &=& \partial_\rho B^{*a\rho}.
\label{DefKT4}
\end{eqnarray}
Furthermore we have,
\bb
\delta x^\mu = 0, \; \delta (dx^\mu)=0.
\ee
The action of $\delta$ is extended to an arbitrary element in
${\cal P}$ by using the rule
\bb
\delta \partial_\mu = \partial_\mu \delta,
\ee
and the fact that $\delta$ is an odd derivation which we take
here to act from the left,
\bb 
\delta (ab)=(\delta a)b+(-)^{\epsilon_{a}}a(\delta b).
\label{parepsi}
\end{equation}
In \ref{parepsi}, $\epsilon_{a}$ is the Grassmann parity of the
(homogeneous) element $a$.  These rules make 
$\delta$ a differential and one has the following
important property \cite{FHST,FH,HenneauxTeitelboim,HenneauxCMP}:
\begin{theorem}
$H_i(\delta)=0$ for $i>0$, where $i$ is the antighost number, i.e,
the cohomology of $\delta$ is empty in antighost number strictly
greater than zero.
\label{propkoszul}
\end{theorem}

One can also show that in degree zero, the 
cohomology of $\delta$ is the algebra of ``on-shell functions"
\cite{FHST,FH,HenneauxTeitelboim,HenneauxCMP}.  
Thus, the Koszul-Tate complex
provides a resolution of that algebra.  For the reader
unaware of the BRST developments, one may view this property as
the motivation for the definitions (\ref{DefKT0}) through
(\ref{DefKT4}).

One has a similar theorem for the cohomology of the exterior
derivative $d$ (for which we also take a left
action, $d(ab)=(da)b+(-)^{\epsilon_a}a(db)$).
\begin{theorem}
The cohomology of $d$ in the algebra of local forms is given by,
\begin{eqnarray}
H^0(d) \simeq R, \\
H^k(d)=0 \hbox{ for } k\not = 0, k\not=n,  \label{dCohomo1} \\
H^n(d) \simeq \hbox{ space of equivalence classes of local forms,}
\label{dCohomo2}
\end{eqnarray}
where $k$ is the form degree and n the spacetime dimension.
In (\ref{dCohomo2}), two local forms are said to be 
equivalent if and only if
they have identical Euler-Lagrange derivatives with respect to all
the fields and the antifields.
\label{poincare}
\end{theorem}
{\bf Proof.} This theorem is known as the algebraic Poincar\'e Lemma.
For various proofs, see
\cite{Vinogr1,Poincare,Brandt,DuboisViolette}.  It should be
mentioned that the theorem holds as such because we allow for an
explicit
$x$-dependence of the local exterior forms (\ref{LocalForm}).  If
the local forms had no explicit $x$-dependence, 
then (\ref{dCohomo1}) would have to be amended as
\bb
H^k(d) \simeq \{\hbox{constant forms} \}
 \hbox{ for } k\not = 0, k\not=n,
\ee
where the constant forms are by definition
the local exterior forms (\ref{LocalForm}) 
with constant coefficients.  We shall denote in the sequel the
algebra of constant forms by $\Lambda^*$ and the subspace of
constants forms of degree $k$ by $\Lambda^k$.  
The following formulation of the Poincar\'e lemma is also
useful.
\begin{theorem}
Let $a$ be a local, closed $k$-form ($k<n$) 
that vanishes when the fields and the antifields are
set equal to zero.  Then, $a$ is $d$-exact.
\label{poincarebis}
\end{theorem}
{\bf Proof.}  The condition that $a$ vanishes when the 
fields and the antifields are
set equal to zero eliminates the constants.   

This form 
of the Poincar\'e lemma holds in  both the algebras
of $x$-de\-pen\-dant and
$x$-independent local exterior forms. 

\section{Characteristic Co\-ho\-mo\-lo\-gy and
Koszul-\newline Tate Complex}
\setcounter{equation}{0}
\setcounter{theorem}{0}

Our analysis of the characteristic cohomology relies upon the
isomorphism established in \cite{BBH1} between 
$H^*_{char}(d)$ and the cohomology $H^*_*(\delta \vert d)$
of $\delta$ modulo $d$.
The cohomology $H^k_i(\delta \vert d)$ in form
degree $k$ and antighost number $i$ is obtained by solving in
the algebra ${\cal P}$ of local exterior forms the equation,
\begin{equation}
\delta a^k_i + db^{k-1}_{i-1}=0,
\end{equation}
and by identifying solutions which differ by $\delta$-exact and
$d$-exact terms, i.e,
\begin{equation}
a^k_i \sim a'^k_i = a^k_i +\delta n^k_{i+1}+dm^{k-1}_i.
\end{equation}
One has
\begin{theorem} \label{CharAnddelta}
\begin{eqnarray}
H^{k}_{char}(d) &\simeq& H^n_{n-k}(\delta \vert d), \; 0<k<n 
\label{CharAndDelta1}\\
\frac{H^{0}_{char}(d)}{R}  &\simeq& H^n_{n}(\delta \vert d).
\label{CharAndDelta2}\\
0 &\simeq& H^n_{n+k}(\delta \vert d),\; k>0
\label{CharAndDelta3}
\end{eqnarray}
\end{theorem}
{\bf Proof.}  Although the proof is standard and can be 
found in \cite{DuboisViolette,BBH1}, we shall repeat 
it explicitly here because it involves 
ingredients which will be needed below.
Let $\alpha$ be a class of $H^k_{char}(d)$ ($k<n$) and let
$a^k_{0}$ be a representative of $\alpha$, $\alpha = [a^k_{0}]$.
One has 
\bb
\delta a^{k+1}_{1} + da^{k}_{0} =0 
\label{MapdDelta1}
\ee
for some
$a^{k+1}_{1}$ since any antifield-independent
form that is zero on-shell
can be written as the $\delta$ of something.  By acting with
$d$ on this equation, one finds that $d a^{k+1}_{1}$
is $\delta$-closed and thus, by  Theorem {\bf \ref{propkoszul}},
that it is $\delta$-exact,
$\delta a^{k+2}_{2}+ da^{k+1}_{1} =0$ for some
$a^{k+2}_{2}$.  One can repeat the procedure until one reaches
degree $n$, the last term $a^n_{n-k}$ fulfilling
\bb
\delta a^n_{n-k} + d a^{n-1}_{n-1-k} = 0, 
\label{MapdDelta2}
\ee
and, of course,
$d  a^n_{n-k}= 0$ (it is a $n$-form).  For future reference we 
collect all the terms appearing in this tower of equations
as
\bb
a^k = a^n_{n-k} + a^{n-1}_{n-1-k} + \dots + a^{k+1}_1 + a^k_0.
\label{Tower}
\ee

The equation (\ref{MapdDelta2}) shows that $a^n_{n-k}$
is a cocycle of the cohomology of $\delta$ modulo $d$, in form-%
degree $n$ and antighost number $n-k$.  Now, given the
cohomological class $\alpha$ of $H^k_{char}(d)$,
it is easy to see, using again 
Theorem {\bf \ref{propkoszul}}, that the corresponding 
element $a^n_{n-k}$ is well-defined in $H^n_{n-k}(\delta \vert d)$.
Consequently, the above procedure
defines an non-ambiguous map $m$ from $H^k_{char}(d)$
to $H^n_{n-k}(\delta \vert d)$.
 
This map is surjective.  Indeed, let $a^n_{n-k}$ be
a cocycle of $H^n_{n-k}(\delta \vert d)$. 
By acting with $d$ on the equation (\ref{MapdDelta2})
and using the second form of the
Poincar\'e lemma (Theorem
{\bf \ref{poincarebis}}), one finds that $a^{n-1}_{n-1-k}$
is also $\delta$-closed modulo $d$. 
Repeating the procedure
all the way down to antighost number zero, one sees that there
exists a cocycle $a^k_0$ of the characteristic 
cohomology such that $m([a^k_0]) = [a^n_{n-k}]$.

The map $m$ is not quite injective, however, because
of the constants.  Assume that $a^k_0$ is mapped on zero.
This means that the corresponding $a^n_{n-k}$ is trivial in
$H^n_{n-k}(\delta \vert d)$, i.e., $a^n_{n-k}
= \delta b^n_{n-k+1} + d b^{n-1}_{n-k}$.  Using the Poincar\'e
lemma (in the second form) one then 
finds successively that $a^{n-1}_{n-k-1}$
$\dots$ up to $a^{k+1}_1$ are all trivial.  The last
term $a^k_0$ fulfills $da^k_0 + \delta db^k_1 = 0$ and
thus, by the Poincar\'e lemma (Theorem {\bf \ref{poincare}}),
 $a^k_0 = \delta b^k_1 + db^{k-1}_0 + c^k$.  In the algebra of
$x$-dependent local forms, the constant
$k$-form
$c^k$ is present only if $k=0$. 
This establishes (\ref{CharAndDelta1}) and (\ref{CharAndDelta2}).
That $H^n_m(\delta \vert d)$ vanishes for $m>n$
is proved in \cite{HenneauxCMP}. 

The proof of the theorem shows also
that (\ref{CharAndDelta1}) holds as
such because one allows for an explicit $x$-dependence of the
local forms.  Otherwise, one must take into account the constant forms
$c^k$ which appear in the analysis of injectivity
and which are no longer exact even when $k>0$, so 
that (\ref{CharAndDelta1}) becomes
\bb
\frac{H^{k}_{char}(d)}{\Lambda^k} \simeq H^n_{n-k}(\delta \vert d),
\ee
while (\ref{CharAndDelta2}) and (\ref{CharAndDelta3}) remain
unchanged.

\section{Characteristic Cohomology and Co\-ho\-mo\-lo\-gy of
$\Delta=\delta+d$}
\setcounter{equation}{0}
\setcounter{theorem}{0}

It is convenient to rewrite the Koszul-Tate differential
in form notations. Denoting the duals with an overline to
avoid confusion with the antifield *-notation, and redefining
the antifields by appropriate multiplicative constants,
one finds that Equations (\ref{DefKT1}) through (\ref{DefKT4}) become
simply
\begin{eqnarray}
\delta {\overline B}^{*a}_1 &+&d{\overline H}^a =0 \nonumber \\
\delta {\overline B}^{*a}_2 &+&d{\overline B}^{*a}_1 =0\nonumber \\
&\vdots& \label{defduaux} \\
\delta {\overline B}^{*a}_{p_a+1}&+&d{\overline B}^{*a}_{p_a} = 0.
\nonumber
\end{eqnarray}
The form ${\overline B}^{*a}_j$ dual to the
antisymmetric tensor density $B^{*a\mu_1 \dots \mu_{p_a+1-j}}$
($j=1,\dots,p_a+1$)
has (i) form degree equal to $n-p_a-1+j$; and (ii) antighost number
equal to $j$.  Since $B^{*a\mu_1 \dots \mu_{p_a+1-j}}$ 
has Grassmann parity $j$
and since the product of $(n-p_a-1+j)$ $dx$'s has Grassmann parity  
$n-p_a-1+j$ ,
each ${\overline B}^{*a}_j$ has same
Grassmann parity $n-p_a-1$ (modulo 2), irrespective of $j$.
This is the same parity as that 
of the $n-p_a-1$-form ${\overline H}^a$ 
dual to the field strengths.

The equation (\ref{defduaux}) can be rewritten as
\bb
\Delta {\tilde H}^a = 0
\ee
with
\bb
\Delta = \delta + d
\ee
and
\bb
{\tilde H}^a = {\overline H}^a + \sum_{j=1}^{p+1} {\overline B}^{*a}_j.
\label{totalB}
\ee
The parity of the exterior form ${\tilde H}^a$ is equal to
$n-p_a-1$.  The regrouping of physical fields with ghost-like
variables is quite standard in BRST theory
\cite{Storaetal}.  Expressions similar (but not
identical) to
(\ref{totalB}) have appeared in the analysis of the Freedman-Townsend
model and of string field theory \cite{Thorn,BauBergSezg}, as
well as in the context of
topological models \cite{Sorella,Baulieu2}.  Note
that for a one-form, expression (\ref{totalB}) reduces
to equation (9.8) of \cite{BBH2}.  Quite generally, it should be
noted that the dual ${\overline H}^a$ to the field
strength $H^a$ is the term of lowest form degree in ${\tilde H}^a$.
It is also the term of lowest antighost number,
namely, zero. At the other end, the term of highest form degree in
${\tilde H}^a$ is ${\overline B}^{*a}_{p_a+1}$, which has form
degree $n$ and antighost number $p_a+1$. If we
call the difference between the form degree and
the antighost number
the ``$\Delta$-degree", all the terms present in
the expansion of ${\tilde H}^a$ have same $\Delta$-degree,
namely $n-p_a-1$.   

The differential $\Delta=\delta+d$ enables
one to reformulate the characteristic cohomology as the cohomology
of $\Delta$. Indeed one has  

\begin{theorem}
The cohomology of $\Delta$ is isomorphic to the
characteristic cohomology,
\begin{eqnarray}
H^k(\Delta) \simeq H^k_{char}(d), \; 0 \leq k \leq n
\end{eqnarray}
where $k$ in $H^k(\Delta)$ is the
$\Delta$-degree, and in $H^k_{char}(d)$ is
the form degree.
\label{ISO}
\end{theorem}
{\bf Proof :}  Let $a^k_0$ ($k<n$) be a cocycle of the characteristic
cohomology. Construct $a^k$ as in the proof of
Theorem {\bf \ref{CharAnddelta}}, formula (\ref{Tower}).
The form $a^k$ is easily seen to be a cocycle of $\Delta$,
$\Delta a^k= 0$, and furthermore, to be uniquely
defined in cohomology given the class of $a^k_0$.
We leave it to the reader to check that the map so defined
is both injective and surjective.  This proves
the theorem for $k<n$.  For $k=n$, the isomorphism
of $H^n(\Delta)$ and $H^n_{char}(d)$ is
even more direct ($da^n_0 = 0 $ is equivalent to
$\Delta a^n_0 = 0 $ and $a^n_0 = db^{n-1}_0 +
\delta b^n_1$ is equivalent to $a^n_0 =
\Delta (b^{n-1}_0 +  b^n_1)$).

Our discussion has also established the following useful
rule:  the term of lowest form degree in a $\Delta$-cocycle
$a$ is a cocycle of the characteristic cohomology.  Its
form degree is equal to the $\Delta$-degree $k$ of
$a$.  For $a={\tilde H}^a$, this reproduce the rule
discussed above theorem \ref{ISO}. Similarly, the term of highest
form degree in
$a$ has always form degree equal to
$n$ if
$a$ is  not a $\Delta$-coboundary (up to a constant),  and
defines an element of $H^n_{n-k}(\delta \vert d)$.
 
Because $\Delta$ is a derivation, its cocycles form an algebra.
Therefore, any polynomial in the
${\tilde H}^a$ 
is also a $\Delta$-cocycle.
Since the form degree is limited by the spacetime dimension
$n$, and since the term ${\overline H}^a$ with
minimum form degree in ${\tilde H}^a$ has form degree
$n-p_a-1$, which is strictly positive,
the algebra generated by the ${\tilde H}^a$
is finite-dimensional.  
We shall show below that these $\Delta$-cocycles are not exact and
that any cocycle of form degree $<n-1$ is a
polynomial in the ${\tilde H}^a$ modulo trivial
terms.  According to the isomorphism expressed by Theorem
{\bf \ref{ISO}}, this is equivalent to proving  Theorem
{\bf \ref{MainResult}}.

\vspace{.3cm}

\noindent
{\bf Remarks: } (i) The $\Delta$-cocycle associated with a
conserved current contains only two terms,
\bb
a=a^n_1 + a^{n-1}_0,
\ee
where $a^{n-1}_0$ is the dual to the conserved current
in question.  The product of such a $\Delta$-cocycle
with a $\Delta$-cocycle of $\Delta$-degree $k$ has
$\Delta$-degree $n-1+k$ and therefore vanishes unless
$k=0$ or $1$.

\noindent
(ii) It will be useful below to introduce another degree $N$ as 
follows. One assigns $N$-degree $0$ to the undifferentiated fields
and $N$-degree $1$ to all the antifields irrespective of their
antighost number.  One then extends the $N$-degree to the 
differentiated variables according to the rule $N(\pp_\mu \Phi) =
N(\Phi) +1$. Thus, $N$ counts the number of derivatives and of
antifields. Explicitly,
\bb
N = \sum_a N_a
\ee
with
\bb
N_a = \sum_J \big[ (|J| \sum_i \partial_{J} B^a_i
\frac{\partial}{\partial_{J} B^a_i}
+ (|J|+1) \sum_\alpha \partial_{J} \phi^{*a}_\alpha
\frac{\partial}{\partial_{J}\phi^{*a}_\alpha} \big].
\label{Na}
\ee
where (i) the sum over $J$ is a sum over all possible derivatives
including the zeroth order one;
(ii) $|J|$ is the differential order of the derivative $\pp_J$
(i.e.,  $|J|=k$ for $\pp_{\mu_1 \dots \mu_k}$); (iii) the sum
over $i$ stands for the sum over the independent components of $B^a$;
and (iv) the sum over $\alpha$ is a sum over the independent
components of all the antifields appearing in the tower associated
with $B^a$ (but there is {\em no} sum over the $p$-form
species $a$ in (\ref{Na})).
The differential $\delta$ 
increases $N$ by one unit.  The 
differentials $d$ and
$\Delta$  have in addition an inhomogeneous piece
not changing the $N$-degree, namely
$dx^\mu (\pp^{explicit} / \pp x^\mu)$, where
$\pp^{explicit} / \pp x^\mu$ sees only
the explicit $x^\mu$-dependence.
The forms ${\tilde H}^a$
have $N$-degree equal to one.  

\section{Acyclicity and Gauge Invariance}
\setcounter{equation}{0}
\setcounter{theorem}{0}

\subsection{Preliminary results}

Under the gauge transformations (\ref{bkgaugeinv}) of the
$p$-form gauge fields, the field strengths and their
derivatives are gauge invariant.  These are the only
invariant objects that can be formed out of the 
``potentials" $B^a_{\mu_1 \dots \mu_{p_a}}$ and their
derivatives.  We shall denote  by ${\cal I}_{Small}$
the algebra of local exterior forms with coefficients
$\omega_{\mu_1 \dots \mu_J}$ that depend only on the field strength
components and their derivatives (and possibly $x^\mu$).  The
algebras ${\cal H}$, ${\cal {\overline H}}$ and ${\cal J}$ respectively
generated by the ($p+1$)-forms $H^a$, ($n-p-1$)-forms
${\overline H}^a$ and ($H^a$, ${\overline H}^a$) are subalgebras
of ${\cal I}_{Small}$.
Since the field equations are gauge invariant and since
$d$ maps ${\cal I}_{Small}$ on ${\cal I}_{Small}$,
one can consider the cohomological problem  
(\ref{cocycleCC}), (\ref{coboundaryCC}) in the algebra 
${\cal I}_{Small}$.  This defines the invariant 
characteristic cohomology $H^{*,inv}_{char}(d)$.

It is natural to
decree that the antifields and their derivatives are also
invariant. This can be more fully justified within the
BRST context, using the property that the gauge transformations
are abelian, but here, it can simply be taken as a useful, consistent
postulate. With these conventions, the differentials $\delta$,
$d$  and $\Delta$ map  the algebra ${\cal I}$ 
of invariant polynomials 
in the field strength components, the antifield components and
their derivatives on itself. Clearly,
${\cal I}_{Small} \subset {\cal I}$. 
The invariant cohomologies $H^{*,inv}(\Delta)$
and  $H^{n,inv}_j(\delta \vert d)$
are defined by considering only local exterior forms that belong
to ${\cal I}$.

In order to analyze the invariant characteristic cohomology 
and to prove the non triviality of the cocycles listed
in Theorem {\bf \ref{MainResult}}, we shall need some
preliminary results on the invariant cohomologies of the
Koszul-Tate differential $\delta$ and of $d$.

The variables generating the algebra
${\cal P}$ of local forms are, together with $x^\mu$ and $dx^\mu$,
$$B_{a\mu_1
\ldots
\mu_{p_a}},
\partial_\rho B_{a\mu_1\ldots \mu_{p_a}},\ldots,B^{*a\mu_1 \ldots
\mu_{p_a-m}}, \partial_\rho B^{*a \mu_1 \ldots
\mu_{p_a-m}},\ldots,B^{*a},\partial_\rho B^{*a},...\ .$$ These
variables can be conveniently split into two subsets. The first
subset of generators will be collectively denoted by the letter
$\chi$. They are given by the field strengths ($H_{a\mu_1 \ldots
\mu_{p_a+1}}$) and their derivatives, the antifields and their
derivatives. The field strengths and their derivatives are not
independent, since they are constrained by the identity $dH^a = 0$
and its differential consequences, but this is not a difficulty for
the considerations of this section.  The
$\chi$'s are invariant under the gauge transformations and they
generate the algebra ${\cal I}$ of invariant
polynomials.
In order to generate the full algebra ${\cal P}$ we need to add to
the $\chi$'s some extra variables that will be
collectively denoted
$\Psi$. The $\Psi$'s contain the field
components $B^{a \mu_1\ldots \mu_{p_a}}$ and their
appropriate derivatives not present in the  $\chi$'s.
The explicit form of the  $\Psi$'s is not needed
here.  All we need to know is that the  $\Psi$'s
are algebraically independent
from the
$\chi$'s and that, in conjunction with
the
$\chi$'s, they generate ${\cal P}$.
\begin{theorem}
Let $a$ be a polynomial in the $\chi$: $a=a(\chi)$. If $a=\delta
b$ then we can choose $b$ such that $b=b(\chi)$.  In particular,
\bb
H^{inv}_j(\delta) \simeq 0 \; \hbox{for } j>0.
\ee
\label{deltachi}
\end{theorem}
{\bf Proof.} We can decompose $b$ into two parts: $b={\overline
b}+ {\overline {\overline b}}$, with ${\overline b} = {\overline
b}(\chi)=b(\Psi=0)$ and ${\overline {\overline b}} =\sum_m
R_m(\chi)S_m(\Psi)$, where $S_m(\Psi)$ contains at least one
$\Psi$. Because $\delta \Psi =0$, we
have, $\delta ({\overline b} + {\overline {\overline b}})=
\delta {\overline b}(\chi) + \sum_m \delta R_m(\chi)S_m(\Psi).$
Furthermore if
$M=M(\chi)$ then $\delta M(\chi) = (\delta M)(\chi).$ We thus
get,
$$a(\chi)=(\delta {\overline b})(\chi) + \sum_m (\delta
R_m)(\chi)S_m(\Psi).$$ The above equation has to be satisfied
for all the values of the $\Psi$'s and in particular for
$\Psi=0$. This means that $a(\chi)=
(\delta {\overline b})(\chi) =\delta {\overline b}(\chi)$.
\begin{theorem}
Let ${\cal H}^k$ be the subspace of form degree $k$
of the finite dimensional algebra
${\cal H}$ of polynomials in the curvature ($p_a+1$)-forms
$H^a$, ${\cal H} = \oplus_k {\cal H}^k$.
One has 
\bb
H^{k,inv}_j(d) = 0, \; k<n, \; j>0
\ee
and
\bb
H^{k,inv}_0(d) = {\cal H}^k, \; k<n. \label{invpoincare2}
\ee
Thus, in particular, 
if $a=a(\chi)$ with $da=0$, antighost $a>0$ and deg $a<n$
then $a=db$ with $b=b(\chi)$.  And if $a$ has antighost number zero,
then $a = P(H^a) +db$, where $P(H^a)$ is a polynomial in the
curvature forms and where $b \in {\cal I}_{Small}$.
\label{invpoincare}
\end{theorem}
{\bf Proof.}  The theorem has been proved in 
\cite{Brandt,Dubois-Violette2} for 1-forms.  It can be extended
straightforwardly to the case of p-forms of odd degree.
The even degree case is slightly different
because the curvatures $(p+1)$-forms $H^a$ are then
anticommuting. It is fully treated in the appendix A.   If
the local forms are not taken  to be explicitly $x$-dependent,
Equation (\ref{invpoincare2}) must be replaced by
\bb
H^{k,inv}_0(d) = (\Lambda \otimes {\cal H})^k.
\ee

\subsection{Gauge invariant $\delta$-boundaries modulo $d$}

We assume in this section that the antisymmetric tensors
$B^a_{\mu_1 \mu_2 \dots \mu_p}$ have all the same degree $p$.
This covers, in particular, the case of a single $p$-form. 
\begin{theorem}
(Valid when the $B^a_{\mu_1 \mu_2 \dots \mu_p}$'s
have all the same form degree $p$).  Let $a_q^n=a_q^n(\chi) \in 
{\cal I}$ be an invariant local $n$-form of
antighost number $q>0$.
If $a_q^n$ is $\delta$-exact modulo
$d$, $a_q^n=\delta \mu_{q+1}^n + d\mu_q^{n-1}$, then
one can assume that $\mu_{q+1}^n$ and $\mu_q^{n-1}$ only depend
on the $\chi$'s, i.e., are invariant ($\mu_{q+1}^n$ and
$\mu_q^{n-1}$ $\in {\cal I}$).
\label{invardelta}
\end{theorem}
{\bf Proof.} The proof goes along exactly the
same lines as the proof of a similar
statement made in \cite{BBH2} (theorem 6.1) for $1$-form gauge fields.
Accordingly, it will not be repeated here\footnote{We
shall just mention a minor point
that has been overlooked in the proof of Theorem 6.1
of \cite{BBH2}, namely, that when $p=1$ in equation
(6.4) of \cite{BBH2}, the form $Z'$ need not vanish (in the
notations of \cite{BBH2}).  However, this does not invalidate
the fact that one can replace $Z'$, $X'_\mu$ etc by invariant
polynomials as the recurrence used in the proof of \cite{BBH2}
and the absence of invariant cohomology for $d$ in form degree
one indicate. This is just what is needed for establishing
the theorem.}. 

\vspace{.1cm}

\noindent
{\bf Remark :} The theorem does not hold if the forms have various
form degrees (see Theorem {\bf \ref{xx}} below).

\section{Characteristic Cohomology for a Single $p$-Form
Gauge Field}
\setcounter{equation}{0}
\setcounter{theorem}{0}

Our strategy for computing the characteristic cohomology
is as follows.  First, we compute  $H^n_*(\delta \vert d)$ 
(cocycle condition, coboundary condition) for a single
$p$-form.  We then use the isomorphism theorems to infer
$H^*_{char}(d)$. Finally, we solve the case of a system involving 
an arbitrary (but finite) number of $p$-forms of various
form degrees.

\subsection{General theorems}
Before we compute $H^n_*(\delta \vert d)$ for a single abelian
$p$-form gauge field $B_{\mu_1 \dots \mu_p}$,
we will recall some general results which will be
needed in the sequel.  These theorems hold for an arbitrary
linear theory of reducibility order $p-1$.

\begin{theorem}
For a linear gauge theory of reducibility order p-1, one has,

\begin{equation}
H^n_j(\delta \vert d)=0,\ \ \ \ j > p+1.
\end{equation}
\label{trivlin}
\end{theorem}
{\bf Proof.} See \cite{BBH1}, Theorem 9.1.  See also 
\cite{Bryant,Vinogr1,Vinogr2}.

Theorem 
{\bf \ref{trivlin}} is particularly useful because it limits the
number of potentially non-vanishing cohomologies. 
The calculation of the characteristic cohomology is further
simplified by the following theorem:
\begin{theorem}
Any solution of  $\delta a +\partial_\rho b^\rho =0$ that is at
least bilinear in the antifields is necessarily trivial.
\label{trivmu}
\end{theorem}
{\bf Proof.} See \cite{BBH1}, Theorem 11.2.

Both theorems hold whether  the local forms
are assumed to have an explicit $x$-dependence
or not.

\subsection{Cocycles of $H^n_{p+1}(\delta \vert d)$}

We have just seen that the first potentially
non-vanishing cohomological group is $H^n_{p+1}(\delta
\vert d)$.  We show
in this section that this group is one-dimensional
and provide  explicit representatives.  
We systematically use the dual notations involving
divergences of antisymmetric tensor densities.

\begin{theorem}
$H^n_{p+1}(\delta \vert d)$ is one-dimensional. One can take as
representatives of the cohomological
classes $a=kB^*$ where $B^*$ is the last antifield, of antighost
number $p+1$ and where $k$ is a 
number.\label{Cohop+1}
\end{theorem}
{\bf Proof.} Any polynomial of antighost number $p+1$ can be
written $a = f B^*+f^\rho\partial_\rho B^*+\ldots+\mu$,
where $f$ does not involve the antifields and where
$\mu$ is at least bilinear in the antifields. 
By adding a divergence to $a$, one can remove the derivatives
of $B^*$, i.e., one can assume $f^\rho = f^{\rho\sigma} = \dots =0$.
The cocycle condition $\delta a +\partial_\rho
b^\rho=0$ implies then $-\partial_\rho f B^{*\rho} 
+\delta \mu +\partial_\rho (b^\rho + f
B^{*\rho})=0$. By taking the Euler-Lagrange derivative of 
this equation with respect to $B^{*\rho}$, one gets
\begin{equation}
-\partial_\rho f + \delta ((-1)^p {\delta^L \mu \over \delta
B^{*\rho}})=0.
\end{equation}
This shows that $f$ is a cocycle of the characteristic cohomology
in degree zero since $\delta$(anything of antighost number one) $
\approx 0$.
Furthermore, if $f$ is trivial in $H^0_{char}(d)$,
then $a$ can be redefined so as to be at least bilinear
in the antifields and thus is also trivial in the cohomology
of $\delta$ modulo $d$. 
Now, the isomorphism
of $H^0_{char}(d)/R$ with $H^n_n(\delta \vert d)$
implies 
$f=k+\delta g$ with $k$ a constant ($H^n_n(\delta \vert d)=0$
because $n>p+1$). As we pointed out, the second term can be
removed by adding a trivial term, 
so we may assume $f=k$.
Writing $a=kB^* +\mu$, we see that $\mu$ has to be a solution
of
$\delta \mu +\partial_\rho b'^\rho=0$ by itself and is therefore
trivial by Theorem {\bf \ref{trivmu}}. So
$H^n_{p+1}(\delta \vert d)$ can indeed be represented by $a=kB^*$.
In form notations, this is just the $n$-form $k {\overline B}^*$.
Note that the calculations are true both in the $x$-dependent and
$x$-independent cases.

To complete the proof of the theorem, it remains to show that
the cocycles $a=kB^*$, which belong to the invariant algebra ${\cal I}$
and which contain the undifferentiated antifields, are non trivial.  
If they were trivial,
one would have according to 
Theorem {\bf \ref{invardelta}},
that ${\overline B}^*  = \delta u + dv$ for some $u$, $v$ also in
${\cal I}$.  But this is impossible, because both $\delta$
and $d$ bring in one derivative of the invariant generators
$\chi$
while ${\overline B}^*$ does not contain derivatives of $\chi$.  
[This derivative counting argument is direct if $u$ and
$v$ do not involve explicitly the spacetime coordinates
$x^\mu$.  If they do, one must expand $u$, $v$ and the equation
${\overline B}^* = \delta u + dv$ according to the number of
derivatives of the fields in order to reach the conclusion.
Explicitly, one sets 
$u = u_0 + \dots + u_k$, $v = v_0 + \dots + v_k$, where $k$
counts the number of derivatives of the $H_{\mu_1 \dots \mu_{p+1}}$
and of the antifields.  The condition ${\overline B}^*=\delta u
+dv$ implies in degree
$k+1$ in the  derivatives that $\delta u_k + d'v_k = 0$, where
$d'$ does not differentiate with respect to the explicit dependence
on $x^\mu$. This relation implies in turn that $u_k$ is
$\delta$-trivial modulo $d'$ since there is no cohomology in
antighost number $p+2$. Thus, one can remove
$u_k$ by adding trivial terms.  Repeating the argument for $u_{k-1}$,
and then for $u_{k-2}$ etc, leads to the desired conclusion.].

\subsection{Cocycles of $H^n_{i}(\delta \vert d)$ with $i \leq p$}

We now solve the cocycle condition for the remaining
degrees. 
First we prove
\begin{theorem}
Let $K$ be the greatest integer such that
$n -K(n-p-1)>1$. 
The cohomological groups $H^n_j(\delta \vert d)$
($j>1$) vanish unless $j = n - k(n-p-1)$,
$k = 1, 2,\dots,K$.  Furthermore, for those
values of $j$,
$H^n_j(\delta \vert d)$
is at most one-dimensional.
\label{Cohop}
\end{theorem}
{\bf Proof.}  We already know that $H^n_j(\delta \vert d)$
is zero for $j>p+1$ and that $H^n_{p+1}(\delta \vert d)$  
is one-dimensional.  Assume thus that the theorem has
been proved for all $j$'s strictly greater than $J<p+1$ and
let us extend it to $J$. In a manner analogous to what
we did in the proof of Theorem {\bf \ref{Cohop+1}}, we
can assume that 
the cocycles of $H^n_J(\delta \vert d)$
take  the form
\bb
f_{\mu_1 \ldots \mu_{p+1-J}}B^{*\mu_1 \ldots \mu_{p+1-J}}+\mu
 \label{particular}
\ee
where $f_{\mu_1 \ldots \mu_{p+1-J}}$ does not
involve the antifields and defines an 
element of $H^{p+1-J}_{char}(d)$.
Furthermore, if $f_{\mu_1 \ldots \mu_{p+1-J}}$ is trivial, then the
cocycle (\ref{particular}) is also trivial. Now, using the
isomorphism $H^{p+1-J}_{char}(d) \simeq H^n_{n-p-1+J}(\delta \vert d)$
($p+1-J >0$),
we see that $f$ is trivial unless $j'=n-p-1+J$, which is strictly
greater than $J$ and is of the form $j' = n - k(n-p-1)$.
In this case, $H^n_{j'}$ is at most one-dimensional. 
Since $J= j' -(n-p-1)=  n -(k+1)(n-p-1)$ is of the
required form, 
the property extends to $J$.  This proves the theorem.

Because we explicitly used the isomorphism 
$H^{p+1-J}_{char}(d) \simeq H^n_{n-p-1+J}(\delta \vert d)$,
which holds only if the local forms are allowed to involve explicitly
the coordinates $x^\mu$, the theorem must be amended for
$x$-independent local forms.  This will be done in section 7.5.

\vspace{.3cm}

Theorem {\bf \ref{Cohop}} 
goes beyond the vanishing theorems of 
\cite{Bryant,Vinogr1,Vinogr2,BBH1}
since it sets further cohomological groups
equal to zero,  in antighost number smaller
than $p+1$.  This is done by viewing the cohomological group
$H^n_i(\delta \vert d)$ as a subset of 
$H^n_{n-p-1+i}(\delta \vert d)$ at a 
{\em higher} value of the antighost number, through
the form (\ref{particular}) of the cocycle
and the isomorphism between $H^{p+1-i}_{char}(d)$
and $H^n_{n-p-1+i}(\delta \vert d)$.  In that manner, the known
zeros at values of the ghost number greater than
$p+1$ are ``propagated" down to values of the ghost number
smaller than $p+1$.
 
To proceed with the analysis, we have to consider two cases:

\noindent
(i) Case I: $n-p-1$ is even. 

\noindent
(ii) Case II: $n-p-1$ is odd. 

\vspace{.1cm}

We start with the simplest case, namely, case I.
In that case, ${\tilde H}$ is a commuting object and we can
consider its various powers $({\tilde H})^k$,
$k=1,2,\dots,K$ with $K$  as in Theorem {\bf \ref{Cohop}}.
These powers have $\Delta$-degree
$k(n-p-1)$. By Theorem {\bf \ref{ISO}},
the term of form
degree $n$ in $({\tilde H})^k$ defines a cocycle of
$H^n_{n-k(n-p-1)}(\delta \vert d)$, which is non trivial as the
same invariance argument used in the previous subsection indicates.
Thus, 
$H^n_{n-k(n-p-1)}(\delta \vert d)$, which
we know is at most one-dimensional,  is
actually exactly one-dimensional and one
may take as representative the term 
of form degree $n$
in $({\tilde H})^k$.  This settles the case when $n-p-1$ is
even. 

In the case when $n-p-1$ is odd, ${\tilde H}$
is an anticommuting object and its
powers $({\tilde H})^k$, $k>0$ all vanish unless $k=1$.
We want to show that $H^n_{n-k(n-p-1)}(\delta \vert d)$ similarly 
vanishes
unless $k=1$.  To that end, it is enough
to prove that $H^n_{n-2(n-p-1)}(\delta \vert d)
 = H^n_{2p+2-n}(\delta \vert d)=0$ 
as the proof of Theorem {\bf \ref{Cohop}} indicates (we
assume, as before, that $2p+2-n>1$ since we only
investigate here the cohomological groups
$H^n_i(\delta \vert d)$ with $i>1$).
Now, as we have seen, the most general cocycle in 
$H^n_{2p+2-n}(\delta \vert d)$
may be assumed to take the form
$a = f_{\mu_{p+2} \ldots \mu_n} B^{*\mu_{p+2} \ldots \mu_n} +\mu$,
where $\mu$ is at least quadratic in the antifields and where
$f_{\mu_{p+2} \ldots \mu_n}$ does not involve the antifields
and defines an element of $H^{n-p-1}_{char}(d)$. But 
$H^{n-p-1}_{char}(d) \simeq H^n_{p+1}(\delta \vert d)$ is
one-dimensional and one may take as representative
of $H^{n-p-1}_{char}(d)$ the dual
$k\epsilon_{\mu_1\ldots \mu_n}H^{\mu_1 \ldots
\mu_{p+1}}$ of the field strength.  
This means that $a$ is necessarily of the form,
\begin{equation}
a=k\epsilon_{\mu_1\ldots \mu_n}H^{\mu_1 \ldots
\mu_{p+1}}B^{*\mu_{p+2} \ldots \mu_n} +\mu,
\label{a}
\end{equation}
and the question to be answered is: for which
value of $k$ can one adjust $\mu$
in (\ref{a}) so that
\begin{equation}
k \epsilon_{\mu_1\ldots \mu_n}H^{\mu_1 \ldots
\mu_{p+1}}\delta B^{*\mu_{p+2} \ldots \mu_n}
+ \delta \mu +\partial_\rho b^\rho =0?
\label{cohomo}
\end{equation}
In (\ref{cohomo}), $\mu$ does not contain $B^{*\mu_{p+2} \ldots 
\mu_n}$
and is at least quadratic in the antifields.  
Without loss of generality, we can assume that it is
exactly quadratic in the antifields and that it
does not contain derivatives, since $\delta$ and $\pp$
are both linear and bring in one derivative. [That
$\mu$ can be assumed to be quadratic is obvious.  That it
can also be assumed not to contain the derivatives
of the antifields is a little
bit less obvious since we allow for explicit $x$-dependence, but
can be easily checked by expanding $\mu$ and $b^\rho$ according
to the number of derivatives of the variables and using
the triviality of the cohomology of $\delta$ modulo
$d$ in the space of local forms that are at least quadratic in
the fields and the antifields].
Thus, we can write
$$\mu = \sigma_{\mu_1 \ldots \mu_n} B^{*\mu_1 \ldots
\mu_{p}}_{(1)} B^{*\mu_{p+1} \ldots \mu_n}_{(2p+1-n)}
+ \mu', $$ where $\mu'$ involves neither $B^{*\mu_{p+2} \ldots \mu_n}
_{(2p+2-n)}$
nor $B^{*\mu_{p+1} \ldots \mu_n}_{(2p+1-n)}$. We have 
explicitly indicated the
antighost number in parentheses in order to keep
track of it.   
Inserting this form of $\mu$ in (\ref{cohomo}) one finds that
$\sigma_{\mu_1
\ldots
\mu_n}$ is equal to $k \epsilon_{\mu_1\ldots \mu_n}$ if 
$2p+1-n>1$ (if $2p+1-n=1$, see below).  One can
then successively eliminate $B^{*\mu_{p} \ldots \mu_n}_{(2p-n)}$,
$B^{*\mu_{p-1} \ldots \mu_n}_{(2p-n-1)}$ etc 
from $\mu$, so that the question ultimately boils down to: is
$$k \epsilon_{\mu_1\ldots \mu_{2j}} B^{*\mu_1 \dots \mu_j}_{(p+1-j)}
\delta
B^{*\mu_{j+1} \dots \mu_{2j}}_{(p+1-j)}$$ ($n$ even $=2j$) or
$$k \epsilon_{\mu_1\ldots \mu_{2j+1}} B^{*\mu_1 \dots 
\mu_{j+1}}_{(p-j)}
\delta
B^{*\mu_{j+2} \dots \mu_{2j+1}}_{(p+1-j)}$$ ($n$ odd $=2j+1$)
$\delta$-exact modulo $d$, i.e., of the form
$\delta \nu + \pp_\rho c^\rho$, where $\nu$
does not involve the antifields $B^*_s$ for
$s>p+1-j$ ($n$ even) or $s>p-j$ ($n$ odd)?
That the answer to this question
is negative unless $k=0$ and $a$ accordingly 
trivial, which is the desired
result, is easily seen by trying to construct
explicitly $\nu$.  We treat for
definiteness the case $n$ even ($n=2j$). One
has $$\nu = \lambda_{\mu_1\ldots \mu_{2j}} 
B^{*\mu_1 \dots \mu_j}_{(p+1-j)}
B^{*\mu_{j+1} \dots \mu_{2j}}_{(p+1-j)}$$ where 
$\lambda_{\mu_1\ldots \mu_{2j}}$ is antisymmetric (respectively
symmetric) for the exchange of $(\mu_1\ldots \mu_{j})$ with
$(\mu_{j+1} \dots \mu_{2j})$ if $j$ is even (respectively
odd) (the $j$-form ${\overline B}^*_{(p+1-j)}$ is odd 
by assumption and this can happen only if the components
$B^{*\mu_1 \dots \mu_j}_{(p+1-j)}$ are odd for $j$ even, or
even for $j$ odd).  From
the equation
\bb
k \epsilon_{\mu_1\ldots \mu_{2j}} B^{*\mu_1 \dots \mu_j}_{(p+1-j)}
\delta
B^{*\mu_{j+1} \dots \mu_{2j}}_{(p+1-j)} = \delta \nu + \pp_\rho c^\rho
\ee
one gets
\bb
k \epsilon_{\mu_1\ldots \mu_{2j}} B^{*\mu_1 \dots \mu_j}_{(p+1-j)}
\pp_\rho B^{* \rho \mu_{j+1} \dots \mu_{2j}}_{(p-j)} =
2 \lambda_{\mu_1\ldots \mu_{2j}} B^{*\mu_1 \dots \mu_j}_{(p+1-j)}
\pp_\rho B^{* \rho \mu_{j+1} \dots \mu_{2j}}_{(p-j)}
+ \pp_\rho c^\rho
\ee
Taking the Euler-Lagrange derivative of this equation
with respect to $B^{*\mu_1 \dots \mu_j}_{(p+1-j)}$ yields next
$$k \big( \epsilon_{\mu_1\ldots \mu_{2j}}-
2 \lambda_{\mu_1\ldots \mu_{2j}} \big)
\pp_\rho B^{* \rho \mu_{j+1} \dots \mu_{2j}}_{(p-j)} = 0,$$ which
implies $k \epsilon_{\mu_1\ldots \mu_{2j}}=
2 \lambda_{\mu_1\ldots \mu_{2j}}$.  This contradicts the symmetry
properties of $\lambda_{\mu_1\ldots \mu_{2j}}$, unless $k=0$,
as we wanted to prove.

\subsection{Characteristic Cohomology}

By means of the isomorphism theorem of section 4, our results on
$H^n_*(\delta \vert d)$ can be translated in terms of the
characteristic cohomology as follows.

\noindent
(i) If $n-p-1$ is odd, the only non-vanishing group of the 
characteristic
cohomology in form degree $<n-1$ is $H^{n-p-1}_{char}(d)$, which is
one-dimensional.  All the other groups vanish.  One may take as
representatives for $H^{n-p-1}_{char}(d)$ the cocycles $k
{\overline H}$.  Similarly, the only non-vanishing group 
$H^{j}(\Delta)$ with
$j<n-1$ is $H^{n-p-1}(\Delta)$ with representatives $k {\tilde H}$
and the only non-vanishing group $H^n_i(\delta \vert d)$ with $i>1$
is
$H^n_{p+1} (\delta \vert d)$ with representatives $k {\overline
B}^*_{p+1}$.

\noindent
(ii) If $n-p-1$ is even, there is further cohomology.
The degrees in which there is non trivial cohomology are
multiples of
$n-p-1$ (considering  again values of  the
form degree strictly smaller than $n-1$).
Thus, there is characteristic
cohomology only in degrees $n-p-1$, $2(n-p-1)$, $3(n-p-1)$ etc.
The corresponding groups are one-dimensional and one may take as
representatives $k{\overline H}$, $k({\overline H})^2$,
$k({\overline H})^3$ etc. There is also non-vanishing
$\Delta$-cohomology for the same values of the
$\Delta$-degree, with representative cocycles given
by $k {\tilde H}$, $k ({\tilde H})^2$, $k ({\tilde H})^3$, etc.
By expanding these cocycles according to the form degree
and keeping the terms of form degree $n$, one gets
representatives for the only non-vanishing groups $H^n_i(\delta \vert 
d)$
( with $i>1$), which are respectively $H^n_{p+1}(\delta \vert d)$,
$H^n_{p+1 -(n-p-1)}$, $H^n_{p+1 -2(n-p-1)}$ etc.

An immediate consequence of our analysis is the following
useful theorem:
\begin{theorem} \label{useful}
If the polynomial $P^k(H)$ of form degree $k<n$ in the
curvature $(p+1)$-form $H$ is $\delta$-exact
modulo $d$ in the invariant algebra ${\cal I}$, then $P^k(H)=0$.
\end{theorem}
{\bf Proof :}  The theorem is straightforward in the
algebra of $x$-independent local forms, as a direct derivative
counting argument shows.  To prove it when an explicit $x$-dependence
is allowed, one proceeds as follows.
If $P^k(H) = \delta a^k_1 + d a^{k-1}_0$ where $a^k_1$ and $a^{k-1}_0
\in {\cal I}$,
then $da^k_1 + \delta a^{k+1}_2 = 0$ for some invariant $a^{k+1}_2$.
Using the results on the cohomology of $\delta$ modulo $d$ that
we have just established, this implies that $a^k_1$
differs from the component of form degree $k$ and antighost
number $1$ of a polynomial $Q({\tilde H})$ by a term of the
form $\delta \rho
+ d \sigma$, where $\rho$ and $\sigma$ are both invariant.
But then, $\delta a^k_1$ has the form $d \big([Q({\tilde H})]^{k-1}_0
+ \delta \sigma \big)$, which implies $P^k(H) = d 
\big(-[Q({\tilde H})]^{k-1}_0
-\delta \sigma + a^{k-1}_0 \big)$, i.e., $P^k(H) = d$(invariant).
According to the theorem on the invariant cohomology of $d$, this
can occur only if $P^k(H) = 0$.

\subsection{Characteristic cohomology in the algebra of 
 $x$-\newline independent local forms}

Let us denote $({\tilde H})^m$ by $P_m$ ($m = 0, \dots, K$).
We have just shown (i) that the most general cocycles of the
$\Delta$-cohomology are given, up to trivial terms,
by the linear combinations $\lambda_m P_m$ with
$\lambda_m$ real or complex numbers; and (ii)
that if $\lambda_m P_m$ is
$\Delta$-exact, then the $\lambda_m$ are all
zero.  In establishing these results, we allowed
for an explicit $x$-dependence of the local
forms (see comments after the proof of
Theorem {\bf \ref{Cohop}}).  How are our results affected if we work
exclusively with local forms with no explicit $x$-dependence?

In the above analysis, it is in calculating the 
cocycles that arise in anti\-ghost number
$<p+1$ that we used the $x$-dependence of the local forms,
through the isomorphism 
$H^{p+1-J}_{char}(d) \simeq H^n_{n-p-1+J}(\delta \vert d)$.
If the local exterior forms are not allowed to depend
explicitly on $x$, one must take the constant $k$-forms
($k>0$) into account.  The derivation goes otherwise
unchanged and one finds that
the cohomology of $\Delta$
in the space of $x$-independent local forms 
is given by the polynomials
in the $P_m$ with coefficients $\lambda_m$ that
are constant forms, $\lambda_m = \lambda_m(dx)$.
In addition, if $\lambda_m P_m$ is $\Delta$-exact,
then, $\lambda_m P_m = 0$ for each $m$.  One cannot
infer from this equation that $\lambda_m$ vanishes, because it is
an exterior form.  One can simply assert that the
components of $\lambda_m$ of form degree $n-m(n-p-1)$ or
lower are zero (when multiplied by $P_m$, the other
components of $\lambda_m$ yields forms of degree $>n$
that identically vanish, no matter what these other
components are).

It will be also useful in the sequel to know the
cohomology of $\Delta'$, where $\Delta'$ is the part of
$\Delta$ that acts only on the fields and antifields,
and not on the explicit $x$-dependence.  One has
$\Delta = \Delta' + d_x$, where 
$d_x \equiv \pp^{explicit}/\pp x^\mu$ sees
only the explicit $x$-dependence.  By the
above result, the cohomology
of $\Delta'$ is clearly given by the polynomials in the
$P_m$ with coefficients $\lambda_m$ that are now
arbitrary spacetime forms, $\lambda_m = \lambda_m(x, dx)$.

\section{Characteristic Cohomology in the General Case}
\setcounter{equation}{0}
\setcounter{theorem}{0}

To compute the cohomology $H^n_i(\delta \vert d)$ 
for an arbitrary set of $p$-forms, 
one proceeds along the lines of the Kunneth theorem.
Let us illustrate explicitly the procedure
for two fields $B^1_{\mu_1 \dots \mu_{p_1}}$  and
$B^2_{\mu_1 \dots \mu_{p_2}}$. One may split
the  differential $\Delta$ as a sum of terms
with definite $N_a$-degrees,
\bb
\Delta = \Delta_1 + \Delta_2 +d_x
\label{split}
\ee
(see  (\ref{Na})).  In (\ref{split}), $d_x$ 
leaves both
$N_1$ and $N_2$ unchanged.  By contrast, $\Delta_1$
increases $N_1$ by one unit without changing $N_2$,
while $\Delta_2$ increases $N_2$  by one unit without changing
$N_1$. The differential $\Delta_1$ acts only on the
fields $B^1$ and its associated antifields (``fields and
antifields of the first set"), whereas the differential $\Delta_2$
acts only on the fields $B^2$ 
and its associated antifields (``fields and
antifields of the second set"). Note that $\Delta_1 +
\Delta_2 = \Delta'$. 

Let $a$ be a cocycle of $\Delta$ with $\Delta$-degree
$< n-1$.  Expand $a$ according to the $N_1$-degree,
\bb
a= a_0 + a_1 + a_2 + \dots + a_m, \; N_1(a_j) = j.
\ee
The equation $\Delta a = 0$ implies $\Delta_1 a_m =0$ for the term
$a_m$ of highest $N_1$-degree. Our
analysis of the $\Delta'$-cohomology for a single $p$-form
yields then $a_m = c_m ({\tilde H}^1)^k + \Delta_1$(something),
where $c_m$ involves only the fields and antifields of the
second set, as well as $dx^\mu$ and
possibly $x^\mu$.
There can be no conserved current in $a_m$ since we assume the
$\Delta$-degree of $a$ - and thus of each $a_j$ - to be strictly
smaller than $n-1$.  Now, the exact term in $a_m$ can be
absorbed by adding to $a_m$ a $\Delta$-exact term, through a
redefinition of $a_{m-1}$.  Once this is done, one finds
that the next equation for $a_m$ and $a_{m-1}$ following
from $\Delta a = 0$ reads
\bb
[(\Delta_2 + d_x) c_m] ({\tilde H}^1)^k + \Delta_1 a_{m-1} = 0.
\ee
But we have seen that $\lambda_m ({\tilde H}^1)^k$ cannot be exact
unless it is zero,
and thus this last equation implies both
\bb
[(\Delta_2 + d_x) c_m] ({\tilde H}^1)^k = 0
\label{vanish}
\ee
and 
\bb
\Delta_1 a_{m-1} = 0.
\ee
Since $({\tilde H}^1)^k$ has independent
form components in degrees
$k(n-p-1)$, $k(n-p-1) +1$ up to degree $n$, we
infer from (\ref{vanish})
that the form components of $(\Delta_2 + d_x) c_m$
of degrees $0$ up to degree $n- k(n-p-1)$ are zero. If we
expand $c_m$ itself according to the form degree,
$c_m = \sum c_{m}^i$, this gives the equations
\bb
\delta c_m^i + d c_m^{i-1} = 0, \; i= 1, \dots, n- k(n-p-1),
\ee
and
\bb
\delta c_m^0 = 0.
\ee
Our analysis of the relationship between the $\Delta$-cohomology
and the cohomology of $\delta$ modulo $d$ indicates then that one
can redefine the terms of form degree $>n- k(n-p-1)$ of $c_m$
in such a way that $\Delta c_m = 0$.
This does not affect the product $c_m ({\tilde H}^1)^k$. 
We shall assume that the 
(irrelevant) higher order terms in $c_m$ have been chosen in that manner.
With that choice, $c_m$ is given, up to trivial terms that can
be reabsorbed, by $\lambda_m ({\tilde H}^2)^l$,
with $\lambda_m$ a number, so that $a_m = \lambda_m
({\tilde H}^2)^l ({\tilde H}^1)^k$ is a $\Delta$-cocycle by itself.
One next repeats successively the analysis for $a_{m-1}$,
$a_{m-2}$ to reach the desired conclusion that
$a$ may indeed be assumed to be a polynomial in the
${\tilde H}^a$'s, as claimed above.

The non-triviality of the polynomials in the ${\tilde H}^a$'s
is also easy to prove.  If $P({\tilde H}) = \Delta \rho$, with
$\rho = \rho_0 + \rho_1 + \dots + \rho_m$, $N_1(\rho_k) = k$, then
one gets at $N_1$-degree $m+1$ the condition $(P({\tilde H}))_{m+1}
= \Delta_1 \rho_m$, which implies $(P({\tilde H}))_{m+1}=0$ and
$\Delta_1 \rho_m=0$, since no polynomial in ${\tilde H}^1$ is
$\Delta_1$-trivial, except zero.  It follows that $\rho_m =
u ({\tilde H}^1)^m$ up to trivial terms that play no role, where
$u$ is a function of the variables of the second set as well as of
$x^\mu$ and $dx^\mu$.  The equation of order $m$ implies then
$(P({\tilde H}))_m = \big( (\Delta_2 + d_x)u \big) ({\tilde H}^1)^m
+ \Delta_1 \rho_{m-1}$.  The non-triviality of the polynomials in
${\tilde H}^1$ in $\Delta_1$-cohomology yields next $\Delta_1 
\rho_{m-1}
=0$ and $(P({\tilde H}))_m = \big( (\Delta_2 + d_x)u \big) 
({\tilde H}^1)^m$.
Since the coefficient of $({\tilde H}^1)^m$ in $(P({\tilde H}))_m$ is
a polynomial in ${\tilde H}^2$, which cannot be 
$(\Delta_2 + d_x)$-exact,
one gets in fact $(P({\tilde H}))_m= 0$ and $(\Delta_2 + d_x)u =0$.
It follows that $\rho_m$ fulfills $\Delta \rho_m=0$ and can be dropped.
The analysis goes on in the same way at the lower values of the
$\Delta_1$-degree, until one reaches the desired conclusion
that the exact polynomial $P({\tilde H})$ indeed vanishes.

In view of the isomorphism between the characteristic cohomology 
and $H^*(\Delta)$, 
this completes the proof of Theorem {\bf \ref{MainResult}}
in the case of two $p$-forms.
The case of more $p$-forms is treated similarly and left to
the reader.

\section{Invariant Characteristic Cohomology}
\setcounter{equation}{0}
\setcounter{theorem}{0}

\subsection{Isomorphism theorems for the invariant
co\-ho\-mo\-lo\-gies} To compute the invariant characteristic
cohomology, we proceed as follows.  First, we establish
isomorphism theorems between $H^{k,inv}_{char}(d)$, 
$H^{n,inv}_{n-k}(\delta \vert d)$ and $H^{k,inv}(\Delta)$.
Then, we compute $H^{k,inv}(\Delta)$ for a single $p$-form.
Finally, we extend the calculation to an arbitrary systems 
of $p$-forms.

\begin{theorem} \label{CharAnddeltaInv}
\begin{eqnarray}
\frac{H^{k,inv}_{char}(d)}{{\cal H}^k} &
\simeq& H^{n,inv}_{n-k}(\delta \vert d), \; 0\leq k<n
\label{CharAndDeltaInv1}\\
0 &\simeq& H^{n,inv}_{n+k}(\delta \vert d),\; k>0
\label{CharAndDeltaInv2}
\end{eqnarray}
\end{theorem}

\begin{theorem}
The invariant cohomology of $\Delta$ is isomorphic to the
invariant characteristic cohomology,
\begin{eqnarray}
H^{k,inv}(\Delta)
\simeq H^{k,inv}_{char}(d), \; 0 \leq k \leq n.
\end{eqnarray}
\label{ISObis}
\end{theorem}

{\bf Proofs :}  First we prove (\ref{CharAndDeltaInv1}).  To
that end we observe that the map $m$ introduced in the
demonstration of Theorem {\bf \ref{CharAnddelta}} maps
$H^{k,inv}_{char}(d)$ on $ H^{n,inv}_{n-k}(\delta \vert d)$.
Indeed, in the expansion (\ref{Tower}) for $a$, all the
terms can be assumed to be invariant on account of Theorem
{\bf \ref{deltachi}}.  The surjectivity of $m$ is also
direct, provided that the polynomials in the
curvature $P(H)$ are not trivial in $H^*(\delta \vert d)$,
which is certainly the case if there is a single $p$-form 
(Theorem {\bf
\ref{useful}}).  We shall thus use Theorem {\bf \ref{CharAnddeltaInv}}
first only in the case of a single $p$-form.  We shall then prove that
Theorem {\bf \ref{useful}} extends to an arbitrary system of
forms of various form degrees, so that the proof  of
Theorem {\bf \ref{CharAnddeltaInv}} will be completed.

To compute the kernel of $m$, consider an element 
$a^k_0 \in {\cal I}$ such that the corresponding
$a^n_{n-k}$ is trivial in  $ H^{n,inv}_{n-k}(\delta \vert d)$.
Then, again as in the proof of Theorem {\bf \ref{CharAnddelta}},
one finds that all the terms in the expansion (\ref{Tower})
are trivial, except perhaps $a^k_0$, which fulfills
$da^k_0 + \delta d b^k_1 = 0$, where $b^k_1 \in {\cal I}$
is the $k$-form appearing in the equation expressing the
triviality of $a^{k+1}_1$, $a^{k+1}_1 =
d b^k_1 + \delta b^{k+1}_2$. This implies
$d(a^k_0 - \delta  b^k_1) = 0$, and thus, by
Theorem {\bf \ref{invpoincare}}, $a^k_0 = P + db^{k-1}_0
+\delta b^k_1$
with $P \in {\cal H}^k$ and $b^{k-1}_0 \in {\cal I}$.
This proves (\ref{CharAndDeltaInv1}), since $P$ is not trivial
in $H^*(\delta \vert d)$ (Theorem {\bf \ref{useful}}). [Again,
we are entitled to use this fact only for a single $p$-form
until we have proved the non-triviality of $P$ in the general
case].

The proof of (\ref{CharAndDeltaInv2}) is a direct consequence of
Theorem {\bf \ref{deltachi}} and parallels step by step
the proof of a similar statement demonstrated for
1-forms in \cite{BBH2} (lemma 6.1).  It will not be
repeated here.  Finally, the proof of Theorem
{\bf \ref{ISObis}}  amounts to observing that the map
$m'$ that sends $[a^k_0]$ on $[a]$ (Equation (\ref{Tower}))
is indeed well defined in cohomology, and
is injective as well as surjective (independently of
whether $P(H)$ is trivial in the invariant cohomology
of $\delta$ modulo $d$).

Note that if the forms do not depend explicitly on $x$, on
must replace (\ref{CharAndDeltaInv1}) by
\bb
\frac{H^{k,inv}_{char}(d)}{(\Lambda \otimes {\cal H})^k}
\simeq H^{n,inv}_{n-k}(\delta \vert d).
\ee

\subsection{Case of a single $p$-form gauge field}
Theorem {\bf \ref{invardelta}} enables one to compute also
the invariant characteristic cohomology for a single
$p$-form gauge field.  Indeed,
this theorem implies that $H^{n,inv}_{n-k}(\delta \vert d)$
and $H^n_{n-k}(\delta \vert d)$ actually coincide since
the cocycles of $H^n_{n-k}(\delta \vert d)$ are invariant
and the coboundary conditions are equivalent.  The isomorphism
of Theorem {\bf \ref{CharAnddeltaInv}} 
shows then that the invariant characteristic
cohomology for a single $p$-form gauge field in form degree $<n-1$
is isomorphic to  the subspace of form degree $<n-1$ of
the direct sum  ${\cal H} \oplus {\cal {\overline H}}$.  Since
the product $H \wedge {\overline H}$ has form degree $n$, 
which exceeds $n-1$, this
is the same as the subspace ${\cal W}$ of Theorem 
{\bf \ref{MainResult2}}. The invariant characteristic cohomology
in form degree $k<n-1$ is thus given by $({\cal H} \otimes
{\cal {\overline H}})^k$, i.e., by the invariant polynomials
in the curvature $H$ and its dual ${\overline H}$ with 
form degree $<n-1$.  Similarly, by the isomorphism of
Theorem {\bf \ref{ISObis}}, the invariant cohomology
$H^{k,inv}(\Delta)$ of $\Delta$ is given by the polynomials
in ${\tilde H}$ and $H$ with $\Delta$-degree smaller 
than $n-1$.

\subsection{Invariant cohomology of $\Delta$ in the general case}
The invariant $\Delta$-cohomology for an arbitrary
system of $p$-form gauge fields follows again from a straightforward
application of the Kunneth formula and is thus
given by the polynomials in the ${\tilde H}^a$'s
and $H^a$'s  with $\Delta$-degree smaller 
than $n-1$.  The explicit proof of this statement works as in
the non-invariant case (for that matter, it is actually more
convenient to use as degrees not $N_1$ and $N_2$, but rather,
degrees counting the number of derivatives of the invariant
variables $\chi$'s. These degrees have the advantage that the
cohomology is entirely in degree zero).  In particular, none of the
polynomials in the ${\tilde H}^a$'s and $H^a$'s is trivial.

The isomorphism of Theorem
{\bf \ref{ISObis}} implies next that the invariant characteristic
cohomology $H^{k,inv}_{char} (d)$ ($k<n-1$)
is given by the polynomials in the 
curvatures $H^a$ and their duals ${\overline H}^a$, restricted to
form degree smaller than $n-1$.  Among these, those
that involve the curvatures $H^a$ are weakly exact, but not
invariantly so. The property of Theorem {\bf \ref{useful}}
thus extends as announced to an arbitrary system of dynamical gauge
forms of various form degrees.

Because the forms have now different form degrees, one may have
elements in $H^{k,inv}_{char} (d)$ ($k<n-1$) that involve
both the curvatures and their duals.  For instance,
if $B^1$ is a $2$-form and $B^2$ is a $4$-form, the
cocycle $H^1 \wedge {\overline H}^2$ is a ($n-2$)-form.
It is trivial in $H^k_{char} (d)$, but not in
$H^{k,inv}_{char} (d)$.

\section{Invariant cohomology of $\delta$ mod $d$}
\setcounter{equation}{0}
\setcounter{theorem}{0}
 
The easiest way to work out explicitly $H^{n,inv}_{n-k}(\delta \vert d)$
in the general case is to use the above isomorphism
theorems, which we are now entitled to do.  Thus, one
starts from $H^{k,inv}(\Delta)$
and one works out the component of form degree $n$ in the
associated cocycles. 

Because one has elements in $H^{k,inv}(\Delta)$
that involves simultaneously both the curvature and its 
$\Delta$-invariant
dual ${\tilde H}$, the property that
$H^{n,inv}_{n-k}(\delta \vert d)$
and $H^n_{n-k}(\delta \vert d)$ coincide may no longer hold.
In the previous example, one would find that $H^{(1)}_{\lambda
\mu \nu} B^{*(2)\lambda \mu \nu}$, which has antighost number
two, is a $\delta$-cocycle modulo $d$, but it cannot
be written invariantly so.
An important case where the isomorphism
$H^{n,inv}_{n-k}(\delta \vert d) \simeq
H^n_{n-k}(\delta \vert d)$ ($k>1$) does hold, however, is when the 
forms
have all the same degrees.

To write down the generalization of Theorem {\bf \ref{invardelta}}
in the case of $p$-forms of different degrees, let
$P(H^a, {\tilde H}^a)$ be a polynomial in the curvatures
$(p_a +1)$-forms $H^a$ and their $\Delta$-invariant
duals ${\tilde H}^a$.  One has $\Delta P = 0$.  We shall be
interested in polynomials of $\Delta$-degree $<n$ that are
of degree $>0$ in both $H^a$ and ${\tilde H}^a$.  The
condition that $P$ be of degree $>0$ in $H^a$ implies that
it is trivial (but not invariantly so), while the condition
that it be of degree $>0$ in ${\tilde H}^a$ guarantees that
when expanded according to the antighost number, $P$ has
non-vanishing components of antighost number $>0$,
\bb
P= \sum_{j=k}^n [P]^j_{j-k}.
\ee
{}From $\Delta P = 0$, one has $\delta [P]^n_{n-k} + d
[P]^{n-1}_{n-k-1} =0$.

There is no polynomial in $H^a$ and ${\tilde H}^a$ with the required
properties if all the antisymmetric tensors $B^a_{\mu_1 \dots 
\mu_{p_a}}$
have the same form degree ($p_a = p$ for all $a$'s) since the 
product $H^a {\tilde H}^b$ has necessarily
$\Delta$-degree $n$.  When there are tensors of different form degrees,
one can construct, however, polynomials $P$ with the given
features.

The analysis of the previous subsection implies straightforwardly.
\begin{theorem} \label{xx}
Let $a_q^n=a_q^n(\chi) \in {\cal I}$
be an invariant local $n$-form of
antighost number $q>0$.
If $a_q^n$ is $\delta$-exact modulo
$d$, $a_q^n=\delta \mu_{q+1}^n + d\mu_q^{n-1}$, then
one has
\bb
a_q^n= [P]_q^n + \delta \mu_{q+1}^{'n} + d\mu_q^{'n-1}
\ee
for some polynomial $P(H^a, {\tilde H}^a)$ of degree
at least one in $H^a$ and at least one in ${\tilde H}^a$, and
where $\mu_{q+1}^{'n}$ and $\mu_q^{'n-1}$ can be assumed to
depend only on the $\chi$'s, i.e.,
to be invariant.
In particular, if all the $p$-form gauge fields have the same
form degree, $[P]_q^n$ is absent and one has
\bb
a_q^n= \delta \mu_{q+1}^{'n} + d\mu_q^{'n-1}
\ee
where one can assume that $\mu_{q+1}^{'n}$ and $\mu_q^{'n-1}$
are invariant ($\mu_{q+1}^{'n}$ and
$\mu_q^{'n-1}$ $\in {\cal I}$).
\end{theorem}

\section{Remarks on Conserved Currents}
\setcounter{equation}{0}
\setcounter{theorem}{0}

That the characteristic cohomology  is
finite-dimensional and  entirely
generated by the duals ${\overline H}^a$'s 
to the field strengths holds only in form degree
$k<n-1$. 
This property is not true
in form degree equal to $n-1$, where there are conserved
currents that cannot be expressed in terms of the forms
${\overline H}^a$,
even up to trivial terms. 

An infinite number of conserved currents that cannot be expressible
in terms of the forms ${\overline H}^a$
are given by
\begin{eqnarray}
T_{\mu \nu \alpha_1 \ldots \alpha_s \beta_1
\ldots \beta_r}= -{1\over 2}({1\over p!}H_{ \mu \rho_1 \ldots
\rho_p, \alpha_1 \ldots \alpha_s} H^{\ \rho_1 \ldots
\rho_p}_{\nu\hspace{0.8cm},\beta_1 \ldots \beta_r}
\nonumber \\ -{1\over (n-p-2)!}H^*_{ \mu \rho_2 \ldots
\rho_{n-p-1}, \alpha_1 \ldots \alpha_s} H^{*\rho_2 \ldots
\rho_{n-p-1}}_{\nu\hspace{1.6cm},\beta_1 \ldots \beta_r}.
\label{gencurr}
\end{eqnarray}
These quantities are easily checked to be conserved
\bb
T^{\mu}_{\nu \alpha_1 \ldots \alpha_s \beta_1\ldots \beta_r ,\mu}
\equiv 0
\ee
and generalize the conserved currents given in
\cite{Lipkin,Morgan,ZZZ} for free electromagnetism.  They are are
symmetric for the exchange of $\mu$ and $\nu$ and are duality
invariant in the critical dimension
$n = 2p+2$ where the field strength and its dual have
same form degree $p+1$.  In this critical dimension, there
are further conserved currents which generalize the ``zilches",
\begin{eqnarray}
Z^{\mu\nu\alpha_1\ldots\alpha_r\beta_1\ldots\beta_s}&=&H^{\mu\sigma_1
\ldots\sigma_p,\alpha_1\ldots\alpha_r}
H_{\ \ \sigma_1\ldots\sigma_p}^{*\nu \hspace{0.9cm} ,\beta_1\ldots
\beta_s} \nonumber \\&&-
H^{*\mu\sigma_1
\ldots\sigma_p,\alpha_1\ldots\alpha_r}
H_{\ \ \sigma_1\ldots\sigma_p}^{\nu \hspace{0.9cm} ,\beta_1\ldots
\beta_s}. 
\end{eqnarray}

Let us prove that the conserved currents (\ref{gencurr}) which contain
an
even total number of derivatives are  not trivial in the space of
$x$-independent local forms. To avoid cumbersome notations we will 
only
look at the currents with no $\beta$ indices.  One may reexpress
(\ref{gencurr}) in terms of the field strengths as
\begin{eqnarray}
T^{\mu \nu \alpha_1 \ldots \alpha_m}&=& -{1\over 2
p!}(H^{ \mu \sigma_1 \ldots
\sigma_p, \alpha_1 \ldots \alpha_m} H^{\nu}_{\ \sigma_1 \ldots
\sigma_p}+
H^{ \mu \sigma_1 \ldots
\sigma_p} H^{\nu\hspace{.8cm}, \alpha_1 \ldots \alpha_m}_{\
\sigma_1 \ldots
\sigma_p})
\nonumber \\
&&+\eta^{\mu\nu}{1\over 2(p+1)!}H_{\sigma_1 \ldots
\sigma_{p+1}}H^{\sigma_1 \ldots
\sigma_{p+1},\alpha_1 \ldots \alpha_m}.
\end{eqnarray}
If one takes the divergence of this expression one gets,
\begin{eqnarray}
T^{\mu \nu \alpha_1 \ldots
\alpha_m}_{\hspace{1.4cm},\mu}&=& \delta K^{\nu \alpha_1 \ldots
\alpha_m}
\end{eqnarray}
where $K^{\nu \alpha_1 \ldots\alpha_m}$ differs from
$k H^{\nu\hspace{.8cm} ,\alpha_1\ldots\alpha_m}_{\
\sigma_1\ldots\sigma_p}B^{*\sigma_1\ldots\sigma_p}$
by a divergence.
It is easy to see that $T^{\mu\nu \alpha_1\ldots\alpha_m}$ is
trivial if and only if
$H^{\nu\hspace{.8cm} ,\alpha_1\ldots\alpha_m}_{\
\sigma_1\ldots\sigma_p}$ $B^{*\sigma_1\ldots\sigma_p}$ is
trivial. So the question is: can we write,
\begin{equation}
H^{\nu\hspace{.8cm} ,\alpha_1\ldots\alpha_m}_{\
\sigma_1\ldots\sigma_p}B^{*\sigma_1\ldots\sigma_p}=\delta
M^{\nu\alpha_1 \ldots \alpha_m}+\partial_\rho N^{\rho
\nu\alpha_1 \ldots \alpha_m}\ 
\end{equation}
for some $M^{\nu\alpha_1 \ldots \alpha_n}$ and
$N^{\rho \nu\alpha_1 \ldots \alpha_m}$?
Without loss of generality, one can assume that $M$ and $N$
have the Lorentz transformation properties indicated
by their indices (the parts of $M$ and $N$ transforming
in other representations would cancel by themselves).  Moreover,
by Theorem {\bf \ref{invardelta}}, one can also assume
that $M$ and $N$ are gauge invariant, i.e., belong to ${\cal I}$.
If one takes into account all the symmetries of the left-hand
side and use the identity $dH=0$, the problem reduces to the
determination of the constant $c$ in,

\begin{eqnarray}
H_{\nu
\sigma_1\ldots\sigma_p,\alpha_1\ldots\alpha_m}B^{*\sigma_1
\ldots\sigma_p}&=&\delta (cH_{\nu \sigma_1 \ldots
\sigma_{p-1}(\alpha_1,\alpha_2\ldots\alpha_m)}B^{*\sigma_1
\ldots \sigma_{p-1}})\nonumber \\ &&\hspace{-1cm}+\partial_\rho
N^{\rho}_{\
\nu\alpha_1 \ldots \alpha_m}+{\hbox{terms that vanish
on-shell}}.
\end{eqnarray}
If one takes the Euler-Lagrange derivative of this equation with
respect to $B^{*\sigma_1\ldots\sigma_p}$ one gets,
\begin{equation}
H_{\nu
\sigma_1\ldots\sigma_p ,\alpha_1\ldots\alpha_m}
\approx (-)^{p+1}cH_{\nu [\sigma_1 \ldots
\sigma_{p-1}\vert(\alpha_1,\alpha_2\ldots\alpha_m)\vert\sigma_p]},
\end{equation}
where the right-hand side is symmetrized in $\alpha_1 \ldots
\alpha_m$ and antisymmetrized in $\sigma_1 \ldots$ $\sigma_p$.
The symmetry properties of the two sides of this equation are not
compatible unless $c=0$. This proves that
$T^{\mu \nu \alpha_1 \ldots \alpha_m}$ (with $m$ even) is not trivial 
in
the algebra of $x$-independent local forms.  It then follows, by
a mere counting of derivative argument, that the
$T^{\mu \nu \alpha_1 \ldots \alpha_m}$ define independent
cohomological classes and cannot be expressed as polynomials
in the undifferentiated dual to the field strengths 
${\overline H}$ with coefficients that are constant forms.

The fact that the conserved currents are not
always expressible in terms of the forms
${\overline H}^a$ makes the validity
of this property for higher
order conservation laws more striking.
In that respect, it should be indicated
that the computation of the characteristic
cohomology in the 
algebra generated by the ${\overline H}^a$
is clearly a trivial question.  The non trivial issue is
to demonstrate that this computation does not miss other cohomological
classes in degree $k<n-1$.

Finally, we point out that the 
conserved currents can all be redefined so as to be
strictly gauge-invariant, apart from a few of them 
whose complete list can be systematically
determined for each given system of $p$-forms.
This point will be fully established in \cite{HKS},
and extends to higher degree antisymmetric tensors a property
established in \cite{BarBrHenn} for one-forms
(see also \cite{Torrebis} in this context).

\section{Introduction of Gauge Invariant Interactions}
\setcounter{equation}{0}
\setcounter{theorem}{0}
The analysis of the characteristic cohomology proceeds in the same
fashion if one adds to the Lagrangian (\ref{Lagrangian}) interactions
that involve higher dimensionality gauge invariant
terms.  As we shall show in \cite{HKS}, these are in general
the only consistent interactions.  These
interactions may increase the derivative order of the fields equations.
The resulting theories should be regarded as effective theories and
can be handled through a systematic perturbation expansion
\cite{Weinberg2}.

The new equations of motion read
\bb
\partial_\mu {\cal L}^{a\mu \mu_1 \mu_2 \dots \mu_{p_a}} = 0
\ee
where ${\cal L}^{a\mu \mu_1 \mu_2 \dots \mu_{p_a}}$ are the 
Euler-Lagrange
derivatives of the Lagrangian with respect to the field strengths
(by gauge invariance, ${\cal L}$ involves only the field strength
components and their derivatives).  These equations can be rewritten 
as 
\bb
d {\overline {\cal L}}^a \approx 0
\ee
where ${\overline {\cal L}}^a$ is the $(n-p_a-1)$-form dual to the
Euler-Lagrange derivatives.

The Euler-Lagrange equations obey the same Noether identities as in
the free case, so that the Koszul-Tate differential takes
the same form, with ${\overline H}^a$ replaced everywhere by
${\overline {\cal L}}^a$.  It then follows that 
\bb
{\tilde {\cal L}}^a = {\overline {\cal L}}^a + 
\sum_{j=1}^{p+1} {\overline B}^{*a}_j
\ee
fulfills
\bb
\Delta {\tilde {\cal L}}^a = 0.
\ee
This implies, in turn, that any polynomial in the
${\tilde {\cal L}}^a$ is $\Delta$-closed. It is also clear that
any polynomial in the ${\overline {\cal L}}^a$ is
weakly $d$-closed.  By making the regularity
assumptions on the higher order terms in the
Lagrangian explained in \cite{BBH1}, one
easily verifies that these are the only cocycles
in form degree $<n-1$,
and that they are non-trivial.  The characteristic
cohomology of the free theory possesses therefore
some amount of ``robustness" since it survives
deformations.  By contrast, the infinite number
of non-trivial conserved currents is not expected
to survive interactions (even gauge-invariant ones).

[In certain dimensions, one may add Chern-Simons
terms to the Lagrangian. These interactions are not strictly
gauge invariant, but only gauge-invariant
up to a surface term.  The equations of motion still
take the form $d$(something) $ \approx 0$, but
now, that ``something" is not gauge invariant.  Accordingly,
with such interactions, some of the cocycles of the characteristic
cohomology are no longer gauge invariant.
These cocycles are removed from the invariant
cohomology, but the discussion proceeds otherwise
almost unchanged and is left to the reader].

\section{Summary of Results and Conclusions}
\setcounter{equation}{0}
\setcounter{theorem}{0}

In this paper, we have completely worked
out the characteristic cohomology $H^k_{char}(d)$
in form degree $k<n-1$ for an arbitrary
collection of free, antisymmetric tensor theories.
We have shown in particular that the
cohomological groups $H^k_{char}(d)$
are finite-dimensional and take a simple form,
in sharp contrast with $H^{n-1}_{char}(d)$,
which  is
infinite-dimensional and appears to be quite
complex.  Thus,  
even though one is dealing with free theories,
which have an infinite number of conserved
local currents,
the existence
of higher degree local conservation laws
is quite constrained.  For instance,
in ten dimensions, there is one 
and only one (non trivial) higher
degree conservation law for
a single 2-, 3-, 4-, 6-, or 8-form
gauge field, in respective form degrees
7, 6, 5, 3 and 1.  It is $d {\overline H} \approx 0$.
For a 5-form, there are two higher degree 
conservation laws, namely
$d {\overline H} \approx 0$ and
$d ({\overline H})^2 \approx 0$, in form degrees
4 and 8.  For a 7-form, there are four  higher degree 
conservation laws, namely
$d {\overline H} \approx 0$, $d ({\overline H})^2 \approx 0$,
$d ({\overline H})^3 \approx 0$ and
$d ({\overline H})^4 \approx 0$,  in form degrees
2, 4, 6 and 8. 

Our results provide at the same
time the complete list of the  
isomorphic groups $H^k(\Delta)$, as well as of $H^n_{n-k}(\delta
\vert d)$.  We have also worked out the
invariant characteristic cohomology,
which is central in the investigation of
the BRST cohomology since it controls the antifield
dependence of BRST cohomological classes \cite{BBH2}.

An interesting feature of the characteristic cohomology in
form degree $<n-1$
is its ``robustness" to the introduction of 
gauge invariant interactions, in contrast to the conserved
currents.

As we pointed out in the introduction,
the characteristic cohomology is interesting for
its own sake since it provides
higher degree local conservation laws.  
But it is also useful in
the analysis of the BRST cohomology.
The consequences of our study will
be fully investigated in a forthcoming
paper \cite{HKS}, where consistent
interactions and anomalies will be studied
(see \cite{Bandell} for the 2-form
case in this context).
In particular, it will be pointed out
how rigid the gauge symmetries are.  We
will also apply our results to compute
the BRST cohomology of the coupled
Yang-Mills-two-form system, where the
field strength of the 2-form is modified
by the addition of the Chern-Simons 3-form
of the Yang-Mills field \cite{ChaplManton}.
This computation will use both the present results and the analysis
of \cite{DV,Brandt,Brandt2,BBH2}.

\section*{Acknowledgments}

M.H. is grateful to LPTHE (Universit\'es Paris VI and
Paris VII) for kind hospitality.  This work has been partly supported
by research funds from F.N.R.S. and a research contract
with the Commission of the European Community.

\appendix

\section{Proof of Theorem \protect\ref{invpoincare}}
 
To prove Theorem {\bf \ref{invpoincare}}, it is convenient
to follow the lines of the BRST formalism.
In that approach, gauge invariance is controlled by the
so-called longitudinal exterior derivative
operator $\gamma$, which acts on the fields and further variables
called ghosts. The construction of $\gamma$ can be found in 
\cite{BV,HenneauxTeitelboim}.  For simplicity, we consider throughout
this appendix the case of a single $p$-form; the general case
is covered by means of the Kunneth formula.
The important point here is the reducibility of the gauge
transformations. Because of this, we need to introduce $p$ ghost
fields:
\begin{equation}
C_{\mu_1 \ldots \mu_{p-1}},\ldots,C_{\mu_1 \ldots
\mu_{p-j}},\ldots, C.
\label{ghosts}
\end{equation}
These ghosts carry a degree called the pure ghost number. The pure 
ghost
number of $C_{\mu_1 \ldots\mu_{p-1}}$ is equal to 1 and increases by
one unit up to $p$ as one moves from the left to the right of
(\ref{ghosts}).
The action of $\gamma$ on the fields and the ghosts is given
by,
\begin{eqnarray}
\gamma B & = & dC_1 \\
\gamma C_1 &=& dC_2 \\
&\vdots& \nonumber \\
\gamma C_{p-1} &=& dC_p \\
\gamma C_p & = &0\\
\gamma (antifield)&=&0.
\end{eqnarray}
In the above equations, $C_{p-j}$ is the $p-j$ form whose
components are $C_{\mu_1 \ldots
\mu_{p-j}}$. For $p$ even, $C_p$ is a commuting object.

One extends $\gamma$ such that it 
is a differential that acts from the left
and anticommutes with $d$.

The motivation behind the above definition is essentially
contained in the following theorem:
\begin{theorem}
The cohomology of $\gamma$ is given by,
\begin{equation}
H(\gamma) = {\cal I} \otimes {\cal C}_p,
\end{equation}
where ${\cal C}_p$ is the algebra generated by the last,
{\em undifferentiated} ghost $C_p$.
In particular, in antighost and pure ghost numbers equal to zero, one
can take as representatives  of the
cohomological class the gauge invariant functions, i.e,
the functions which depend solely on the field strengths and their
derivatives. [It is in that sense that the differential $\gamma$
incorporates gauge invariance].
\end{theorem}
The proof of this theorem follows the lines given in
\cite{Dubois-Violette2}, by redefining the generators
of the algebra so that $\gamma$ takes the standard form
$\gamma x_i = y_i$, $\gamma y_i = 0$, $\gamma z_\alpha =0$
in terms of the new generators $x_i, y_i, z_\alpha$.  The
paired variables $x_i, y_i$ disappear from the cohomology,
which is entirely generated by the unpaired variables
$z_\alpha$.  In the present case, one easily convince
oneself that the generators of ${\cal I} \otimes {\cal C}_p$
are precisely of the $z_\alpha$-type, while the other
generators come in pairs.
The derivatives of the last ghost $C_p$ are paired with the
symmetrized derivatives of the next-to-last ghost $C_{ p-1 \mu}$,
the other derivatives of the next-to-last ghost $C_{p-1 \mu}$
which may be expressed as derivatives of the ``curvatures"
$\partial_\mu C_{p-1 \nu} - \partial_\nu C_{p-1 \mu}$, are
paired with the derivatives of the previous ghost
$\partial_{\alpha_1 \dots \alpha_k} C_{(p-2),\mu \nu}$ involving
a symmetrization, say on $\alpha_1$ and $\mu$ etc. The details
present no difficulty and are left to the reader.

According to the theorem,
any solution of the equation $\gamma a =0$ can be written,
\begin{equation}
a=\sum_l \alpha_l(\chi)C^l +\gamma b.
\end{equation}
Furthermore, if $a$ is $\gamma$-exact, then one has $\alpha_l \equiv
0$ since the various powers of $C$ are linearly independent.

The previous theorem holds independently of whether $p$ is
even or odd.  We now assume that $p$ is even, so that the curvature
$(p+1)$-form $H$ is anticommuting and the last ghost $C_p$ is
commuting, and prove Theorem {\bf \ref{invpoincare}} in that case
(the case when $p$ is odd parallels the $1$-form case and so 
need not be treated here). 

Assume that $da^k_0=0$ with $a_0^k$ a
polynomial in the field strengths and their derivatives. By the
Poincar\'e lemma we have $a^k_0=da^{k-1}_0$, but there is no guarantee
that $a^{k-1}_0$ is also in ${\cal I}_{small}$. Acting with $\gamma
$ on this equation we get, again using the Poincar\'e lemma, 
$\gamma a^{k-1}_0+d a^{k-2}_1=0$. One can thus construct a tower
of equations which take the form,
\begin{eqnarray}
a^k_0&=&da^{k-1}_0 \\
\gamma a^{k-1}_0+d a^{k-2}_1&=&0 \\
&\vdots& \nonumber \\
\gamma a^{k-1-q}_q+d a^{k-2-q}_{q+1}&=&0 \label{eqsubst}\\
\gamma a^{k-2-q}_{q+1}&=&0 .
\end{eqnarray}
Let $r=k-2-q$ and $q+1=m$. If $m=pl$ then the last equation of the
tower implies,
\begin{equation}
a^r_m=C_p^l P  +\gamma a^r_{m-1},
\label{amr}
\end{equation}
with $P \in {\cal I}_{small}$.

If $m \not = pl$ then we simply have $a^r_m=\gamma a^r_{m-1}$. In
that case, an allowed redefinition of the tower allows one to
suppose that the tower stops earlier with $\gamma a^r_{m'} =0$ and
$m'=pl$. An allowed redefinition of the tower simply adds to $a_0^k$
a term of the form $db^{k-1}_0$ where $b^{k-1}_0$ is gauge
invariant.

So from now, we shall assume that indeed $m=pl$. If we substitute
(\ref{amr}) in (\ref{eqsubst}) we get, 
\begin{equation}
\gamma (a^{r+1}_{m-1} + lC_{p-1}C_p^{l-1})+C^l_p dP=0.
\end{equation}
(the trivial term $\gamma a^r_{m-1}$ is absorbed in an
allowed redefinition of the tower). Since the action of
$d$ is well defined in ${\cal I}_{small}$ this implies $dP=0$. The
form degree of $P$ is strictly smaller than the form degree of
$a_0^k$ so let us make the recurrence hypothesis that the theorem
holds for
$P$. Because we treat the case $p$ even, $H$ is odd and
$P=c'H+c+dQ$ where $c$ and $c'$ are constants and $Q \in {\cal
I}_{small}$. We thus have,
\begin{equation}
a^r_m=cC^l_p  +c'C_p^l H+ dQ C^l_p.
\label{formeamr}
\end{equation}
The last two terms in (\ref{formeamr}) are trivial. For the first
one we have,
\begin{equation}
dBC^l_p=d(QC^l_p)-\gamma (QlC_{p-1}C_p^l).
\end{equation}
Then we note that,
\begin{eqnarray}
HC_p^l&=&{1\over l+1}(d({\sum_{i_1+\ldots +i_{l+1}=pl \atop 0\leq
i_1 \leq p,\ldots,0\leq i_{l+1}\leq p}
C_{i_1}C_{i_2}\ldots C_{i_{l+1}}})\\ &&\hspace{1.5cm}-\gamma 
(\sum_{i_1+\ldots +i_{l+1}=pl-1 \atop 0\leq
i_1 \leq p,\ldots,0\leq i_{l+1}\leq p}
C_{i_1}C_{i_2}\ldots C_{i_{l+1}})).
\end{eqnarray}
To prove the above identity one sets $C_0 =B$ and $C_{p+1}\equiv
0$ since the
ghost of highest pureghost is $C_p$. One also uses the fact that
$\gamma C_i =dC_{i+1}$.

This shows that we only
need to consider the bottom $a^r_m=cC^l_p$. Let us now prove that if
$l>1$ then
$c=0$. We can write,
\begin{equation}
a^{pl}_0=c{\sum_{i_1+\ldots +i_{l}=pl \atop 0\leq
i_1 \leq p,\ldots,0\leq i_{l+1}\leq p}}C_{i_1}C_{i_2}\ldots
C_{i_{l}}.
\end{equation}
It is easy to show that for $(p-1)l +1\leq k \leq pl$ we have,
\begin{equation}
d({\sum_{i_1+\ldots +i_{l}=k \atop 0\leq
i_1 \leq p,\ldots,0\leq i_{l}\leq p}
C_{i_1}C_{i_2}\ldots C_{i_{l}}})=\gamma 
(\sum_{i_1+\ldots +i_{l}=k-1 \atop 0\leq
i_1 \leq p,\ldots,0\leq i_{l}\leq p}
C_{i_1}C_{i_2}\ldots C_{i_{l}}).
\end{equation}
This implies that,
\begin{equation}
a^{p(l-1)}_{p}=(-)^pc{\sum_{i_1+\ldots +i_{l}=p(l-1) \atop 0\leq
i_1 \leq p,\ldots,0\leq i_{l+1}\leq p}}C_{i_1}C_{i_2}\ldots
C_{i_{l}}+C_p^{l-1}Q,
\end{equation}
where $Q \in {\cal I}_{small}$. If we substitute this in $\gamma 
a^{p(l-1)-1}_{p+1}+da^{p(l-1)}_{p}=0$ we get,
\begin{eqnarray}
(-1)^pclHC_p^{l-1}+dQ C_p^{l-1}+\gamma
((-)^pc{\sum_{i_1+\ldots +i_{l}=p(l-1)-1 \atop 0\leq i_1 \leq
p,\ldots,0\leq i_{l+1}\leq p}}C_{i_1}C_{i_2}\ldots
C_{i_{l}}\nonumber\\
+Q(l-1)C_{p-1}C_p^{l-2}+a^{p(l-1)-1}_{m+p+1})=0.
\end{eqnarray} The above equation implies,
\begin{equation}
dQ+(-1)^pclH=0.
\end{equation}
Because the form degree of $H$ is strictly smaller than the form 
degree of
$a_0^k$, the above recurrence hypothesis tells us that this equation
is impossible unless $c=0$.

To show that $c=0$ we had to lift the bottom $a^{pl}_0$ $p+1$
times. This is only possible when the tower has $p+1$ steps, which 
is the case when $l >1$. If $l=1$ then the bottom is $a^l_0=c
C_p$. This bottom can be lifted to the top of the tower and
yields $a_0^{p+1}=cH+dN$, $N \in {\cal I}_{small}$. 

To validate
the recurrence hypothesis we observe that if $a_0^k$ is
of form degree 0 then necessarily $a^0_0=k$. 

This ends the proof of the theorem and
shows that for
$p$ even we have $a=c+c'H+dN$, $N \in {\cal I}_{small}$,
as desired.

\end{document}